\documentclass[fleqn,usenatbib]{mnras}

\usepackage{newtxtext,newtxmath}

\usepackage[T1]{fontenc}

\DeclareRobustCommand{\VAN}[3]{#2}
\let\VANthebibliography\thebibliography
\def\thebibliography{\DeclareRobustCommand{\VAN}[3]{##3}\VANthebibliography}

\usepackage{graphicx}	
\usepackage{amsmath}
\usepackage{amsfonts,amssymb}

\usepackage{tikz,xcolor,hyperref}

\definecolor{lime}{HTML}{A6CE39}
\DeclareRobustCommand{\orcidicon}{%
	\begin{tikzpicture}
	\draw[lime, fill=lime] (0,0) 
	circle [radius=0.16] 
	node[white] {{\fontfamily{qag}\selectfont \tiny ID}};
	\draw[white, fill=white] (-0.0625,0.095) 
	circle [radius=0.007];
	\end{tikzpicture}
	\hspace{-2mm}
}

\foreach \x in {A, ..., Z}{%
	\expandafter\xdef\csname orcid\x\endcsname{\noexpand\href{https://orcid.org/\csname orcidauthor\x\endcsname}{\noexpand\orcidicon}}
}

\title[Tidal dissipation in convective stars or planets]{Efficiency of tidal dissipation in slowly rotating fully convective stars or planets}

\author[J. Vidal \& A. J. Barker]{
J\'er\'emie Vidal\orcidA{}\thanks{E-mail: vidalje63@gmail.com}$^{1}$,
Adrian J. Barker\orcidB{}$^{1}$
\\
$^{1}$Department of Applied Mathematics, School of Mathematics, University of Leeds, Leeds, LS2 9JT, UK
}

\date{Accepted  2020 July 23. Received 2020 July 23; in original form 2020 July 6}

\pubyear{2020}

\begin{document}
\label{firstpage}
\pagerange{\pageref{firstpage}--\pageref{lastpage}}
\maketitle

\newcommand{\wcv}{\omega_{cv}}

\begin{abstract}
Turbulent convection is thought to act as an effective viscosity in damping equilibrium tidal flows, driving spin and orbital evolution in close convective binary systems. Compared to mixing-length predictions, this viscosity ought to be reduced when the tidal frequency $|\omega_t|$ exceeds the turnover frequency $\wcv$ of the dominant convective eddies, but the efficiency of this reduction has been disputed. We reexamine this long-standing controversy using direct numerical simulations of an idealized global model. We simulate thermal convection in a full sphere, and externally forced by the equilibrium tidal flow, to measure the effective viscosity $\nu_E$ acting on the tidal flow when $|\omega_t|/\wcv \gtrsim 1$. We demonstrate that the frequency reduction of $\nu_E$ is correlated with the frequency spectrum of the (unperturbed) convection. For intermediate frequencies below those in the turbulent cascade ($|\omega_t|/\wcv \sim 1-5$), the frequency spectrum displays an anomalous $1/\omega^\alpha$ power law that is responsible for the frequency-reduction $\nu_E \propto 1/|\omega_t|^{\alpha}$, where $\alpha < 1$ depends on the model parameters. We then get $|\nu_E| \propto 1/|\omega_t|^{\delta}$ with $\delta > 1$ for higher frequencies, and $\delta=2$ is obtained for a Kolmogorov turbulent cascade. A generic $|\nu_E| \propto 1/|\omega_t|^{2}$ suppression is next found for higher frequencies within the dissipation range of the convection (but with negative values). Our results indicate that a better knowledge of the frequency spectrum of convection is necessary to accurately predict the efficiency of tidal dissipation in stars and planets resulting from this mechanism.
\end{abstract}


\section{Introduction}
\label{sec:intro}
Turbulent convection in stars is believed to dissipate the tidal shear excited by gravitational interactions in close stellar binary or planetary systems, and this process can play an important role in determining the orbital and spin evolution of low-mass binary stars or short-period planets \citep[e.g.][]{Mazeh2008,ogilvie2014tidal}. The time-scale for these evolutionary processes is inversely proportional to the effective viscosity, and so estimating the stellar (or planetary) viscosity is of crucial importance in applications. 
The laminar viscosity in convective envelopes is much too small to be relevant for tidal evolution \citep[e.g.~in the Sun;][]{hanasoge2014quest}, and so turbulent convection is usually thought to act as an effective turbulent viscosity $\nu_E$ that is responsible for damping oscillatory tidal flows. 
This mechanism is usually invoked to explain the circularization and synchronization of binary systems containing low-mass or solar-like main-sequence stars  \citep[e.g.][]{zahn1989tidal,zahn1989tidalb,meibom2005robust,meibom2006observational,van2016orbital,lurie2017tidal,triaud2017eblm,von2019eblm}, and evolved stars \citep[e.g.][]{verbunt1995tidal,Beck2018,price2018binary}. 

The effective viscosity due to convection can be estimated by neglecting the oscillatory nature of the tidal flow such that $\nu_E \simeq \nu_{cv}$ \citep[leading to the standard constant lag-time tidal model, e.g.][]{Alexander1973,Eggleton1998}, where $\nu_{cv}$ is the turbulent viscosity predicted by mixing-length theory \citep[MLT, e.g.][]{Spiegel1971}. Understanding and characterizing the interaction between oscillatory tidal flows and turbulent convection has been referred to as the Achilles' heel of tidal theory \citep{zahn2008tidal}. \citet{zahn1966marees} first realized that $\nu_{E}$ ought to be reduced when the tidal frequency $|\omega_t|$ is faster than the turnover frequency $\wcv$ of the dominant convective eddies.
The magnitude of this inhibition has been however disputed \citep[e.g.][]{goodman1997fast}, and two contradictory prescriptions have been used. 
\citet{zahn1966marees,zahn1989tidal} proposed the linear scaling 
\begin{equation}
	\nu_E \propto \nu_{cv} \, (|\omega_t| / \wcv)^{-1},
	\label{eq:reductionLIN} 
\end{equation}
which is derived by applying MLT arguments assuming that the largest eddies dominate the dissipation, but \citet{goldreich1977turbulent}
proposed instead a quadratic reduction
\begin{equation}
    \nu_E \propto \nu_{cv} \, (|\omega_t| / \wcv)^{-2}
    \label{eq:reductionQUA}
\end{equation}
that is derived by assuming that the dominant contribution to the effective viscosity at short tidal periods comes from eddies in the turbulent (Kolmogorov) cascade with a turnover time-scale comparable with the oscillation period. 

When equations (\ref{eq:reductionLIN})-(\ref{eq:reductionQUA}) are evaluated in stellar models, they typically lead to very different predictions for tidal evolutionary time-scales \citep[e.g.][]{price2018binary}. Thus, application of tidal theory to convection zones remains uncertain, and determining the correct frequency-reduction law of the turbulent viscosity is crucial before we can apply tidal theory to interpret observations of close binaries \citep[e.g.][]{kirk2016kepler,lurie2017tidal,van2016orbital,triaud2017eblm,price2018binary} and possibly also short-period planetary orbits \citep[e.g.][]{Rasio1996}. 
It is possible that the two laws could be valid in different frequency ranges.
Indeed, scaling (\ref{eq:reductionLIN}) seems to work well when applied to certain stellar oscillations \citep{gonczi1982local} or in early calculations of pre-main sequence circularization \citep{zahn1989tidalb}, whereas quadratic scaling (\ref{eq:reductionQUA}) could be relevant for much shorter forcing periods, such as those that are relevant for the interaction between acoustic modes and convection \citep{goldreich1977solar,goldreich1994excitation,Samadi2001}. 

The frequency-reduction law of the turbulent viscosity acting on tidal flows has been also independently revisited with direct numerical simulations (DNS). 
The two laws were first recovered in separate studies, which support either the linear scaling \citep{penev2007dissipation,penev2009direct}
or the quadratic suppression \citep{ogilvie2012interaction,braviner2016stellar,duguid2020tidal}.
The coexistence of the two scaling laws has however been found subsequently, using an idealized turbulence model \citep{goldman2008effective} and in our previous global DNS \citep{vidal2020turbulent}.
These recent results have the potential to reconcile the previous theoretical and numerical findings. 
Moreover, the recent numerical findings have shed light on the fact that the two scaling laws may be appropriate for different reasons than those originally suggested.
On the one hand, the quadratic suppression has been convincingly found for high frequencies $|\omega_t|\gg \wcv$, particularly those outside the turbulent cascade 
\citep{ogilvie2012interaction,braviner2016stellar,duguid2020tidal,vidal2020turbulent}. 
On the other hand, the linear reduction, which has been only observed in an intermediate-frequency range (with $|\omega_t|\sim \wcv$), may be correlated with the frequency spectrum of the (unperturbed) convection. 
Indeed, the convective frequency spectrum is expected to be flatter than the Kolmogorov frequency spectrum in that range, as reported for Boussinesq \citep{vidal2020turbulent} or compressible \citep[e.g.][]{penev2011three,horst2020fully} convection, such that predictions (\ref{eq:reductionLIN})-(\ref{eq:reductionQUA}) may not be generic. 

Owing to the importance of this problem to understand tidal evolution, we continue our numerical investigation \citep{vidal2020turbulent} using global DNS of convection in the presence of the equilibrium tidal flow to gain robust physical insights into the efficiency of tidal dissipation in slowly rotating convective stars or planets.
Our global model complements the previous local studies in Cartesian geometry \citep[e.g.][]{ogilvie2012interaction,braviner2016stellar,duguid2020tidal}, in that we study more realistic tidal flows, and we explore convective flows in stellar-like (or planetary-like) spherical domains in which the flow is free from the influence of artificial periodic (or shearing-periodic) boundary conditions. On the other hand, global DNS are typically more computationally-demanding than local DNS, which prevents us from studying very long tidal periods relative to convective time-scales.

The paper is organized as follows. 
We present our global model and numerical methods in Section \ref{sec:model}, and discuss the general properties of the unperturbed convection in Section \ref{sec:conv}. 
Direct computations of the turbulent viscosity are presented in Section \ref{sec:results}. 
The implications of our results are presented in Section \ref{sec:discussion}, and we conclude the paper in Section \ref{sec:ccl}.

\section{Formulation of the problem}
\label{sec:model}
\subsection{Convection model}
We study the interplay between tidal flows and convection using an idealized model of fully convective stars or giant planets.
We model a full sphere of radius $R$ and volume $V$, filled with a fluid of uniform (laminar) kinematic viscosity $\nu$ and thermal diffusivity $\kappa$, and employ spherical coordinates $(r,\theta,\phi)$ centered on the body. 
The body possibly rotates at the angular velocity $\Omega_s \boldsymbol{1}_z$, where $\boldsymbol{1}_z$ is the Cartesian unit vector along the polar axis. 
We model convection in the Boussinesq approximation \citep{Spiegel1971}, considering slight fluctuations of temperature $\Theta$ and velocity from the motionless conduction state $T_0(r)$ sustained by the homogeneous internal heating source $\mathcal{Q}_T$. 
The gravitational field is $\boldsymbol{g} = - \gamma\, \boldsymbol{r}$, where $\boldsymbol{r}$ is the position vector and $\gamma$ is a constant, which represents the leading-order component for a low-mass body that is not very centrally condensed. 
The primary body is also subjected to tidal forcing from an orbiting companion, which drives large-scale tidal flows in the fluid interior \citep{ogilvie2014tidal,le2015flows}.
Following \citet{goodman1997fast}, we divide the total velocity field $\boldsymbol{u}+\boldsymbol{U}_0$ into two components, a turbulent convective flow $\boldsymbol{u}$ and a background large-scale tidal flow $\boldsymbol{U}_0$ (see below). 

We employ dimensionless quantities for the simulations, adopting $R$ as the length scale, the viscous time-scale ${R^2}/{\nu}$ as the time-scale, and $(\nu \mathcal{Q}_T R^2)/(6 {\kappa}^2)$ as the unit of temperature \citep[as in][]{vidal2020turbulent}. 
The dimensionless Boussinesq equations for the fluctuations $[\boldsymbol{u},\Theta]$ in the rotating frame are
\begin{subequations}
\allowdisplaybreaks
  \label{eqModeling:GeneralEquations}
	\begin{align}
    \frac{\partial \boldsymbol{u}}{\partial t} + (\boldsymbol{u} \boldsymbol{\cdot} \boldsymbol{\nabla} ) \,  \boldsymbol{u} &= 
    - \boldsymbol{\nabla} p +  \boldsymbol{\nabla}^2 \boldsymbol{u} + Ra \, \Theta \, \boldsymbol{r} - \boldsymbol{f}, \\
    \frac{\partial \Theta}{\partial t} + (\boldsymbol{u} \boldsymbol{\cdot} \boldsymbol{\nabla}) \, \Theta &=  \frac{1}{Pr} \left [ 2 \, \boldsymbol{u} \boldsymbol{\cdot} \boldsymbol{r} + \nabla^2 \Theta \right ] - \mathcal{Q}, \\
    \boldsymbol{\nabla}\boldsymbol{\cdot} \boldsymbol{u} &= 0,
    \end{align}
\end{subequations}
with the dimensionless (reduced) pressure $p$ and
\begin{subequations}
\begin{align}
\boldsymbol{f} &= (2/E) \, \boldsymbol{1}_z \times \boldsymbol{u} + ( \boldsymbol{u} \boldsymbol{\cdot} \boldsymbol{\nabla} ) \, \boldsymbol{U}_0 + ( \boldsymbol{U}_0 \boldsymbol{\cdot} \boldsymbol{\nabla} ) \, \boldsymbol{u}, \\
\mathcal{Q} &= (\boldsymbol{U}_0 \boldsymbol{\cdot} \boldsymbol{\nabla}) \, \Theta.
\end{align}
\end{subequations}
We have discarded the term $(\boldsymbol{U}_0 \boldsymbol{\cdot} \boldsymbol{\nabla}) \, T_0$ in the temperature equation, since it should be negligible when $\beta \ll 1$ \citep[e.g.][in the ellipsoidal geometry]{lai1993ellipsoidal}. 
We have also introduced in equations (\ref{eqModeling:GeneralEquations}) the Rayleigh number $Ra$, the Prandtl number $Pr$ and the Ekman number $E$. 
They are given by
\begin{subequations}
\begin{equation}
    Ra = \frac{\alpha_T \gamma \mathcal{Q}_T R^6}{6\nu \kappa^2}, \quad Pr = \frac{\nu}{\kappa}, \quad E = \frac{\nu}{\Omega_s R^2},
    \tag{\theequation a--c}
\end{equation}
\end{subequations}
where $\alpha_T$ is the thermal expansion coefficient. The Rayleigh number measures the strength of the convective driving, and the Ekman number the strength of viscous diffusion with respect to global rotation. Since many low-mass stars are slow rotators \citep[e.g.][]{nielsen2013rotation,Newton2018},  we will mainly ignore global rotation in the DNS by setting $E=+\infty$ (though we will also consider a few slowly rotating cases, see below).

Equations (\ref{eqModeling:GeneralEquations}) are complemented with boundary conditions at the (dimensionless) spherical boundary $r=1$. 
For the temperature, we employ the isothermal condition $\Theta=0$ (we expect to obtain similar results using fixed flux conditions). 
To avoid spurious numerical issues associated with angular momentum conservation in global simulations of tidal flows \citep[e.g. as observed in][]{favier2014non}, we enforce the no-slip (NS) boundary conditions (BC) $\boldsymbol{u}=\boldsymbol{0}$. 
The latter BC does not qualitatively affect the (small-scale) turbulent flows driven in the bulk in our simulations, compared to the more realistic stress-free (or free-surface) BC for stellar applications.

\subsection{Tidal forcing}
\begin{figure}
    \centering
    \includegraphics[trim={0cm 0cm 0cm 0cm}, clip,,width=0.45\textwidth]{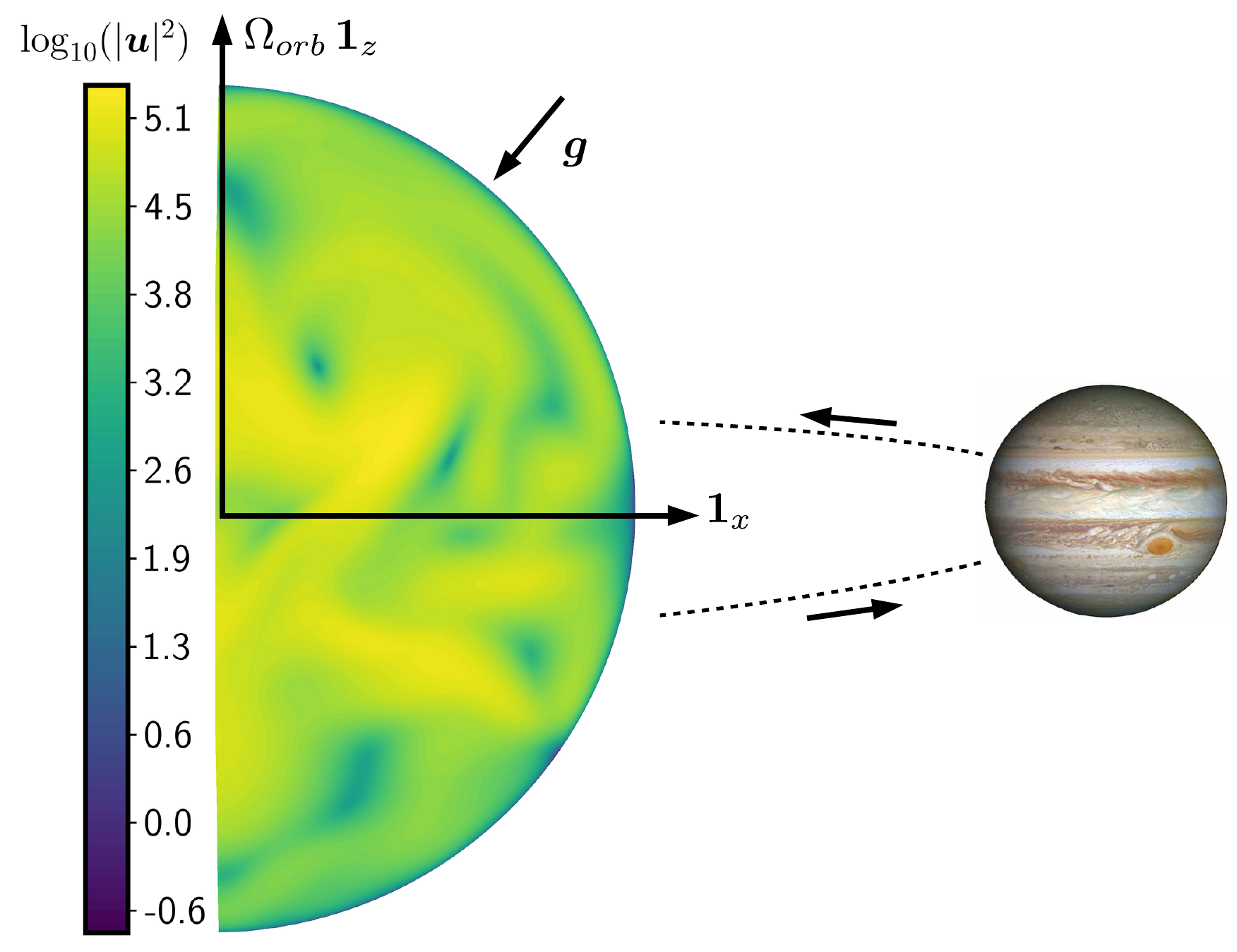}
    \caption{Sketch of the tidal problem in the inertial frame.
    The companion is orbiting around the fluid body in the orbital plane (dashed line), with angular velocity $\Omega_{orb} \, \boldsymbol{1}_z$. Color bar illustrates $\log_{10}(|\boldsymbol{u}|^2)$ for our DNS with $Ra=8 \times 10^6$.} 
    \label{fig:sketch}
\end{figure}

Previous numerical studies modeled the tidal flow with either an (ad-hoc) external forcing \citep{penev2009direct}, or with a background unidirectional shear flow in a shearing box \citep{ogilvie2012interaction,braviner2016stellar,duguid2020tidal}. 
For a more realistic astrophysical model, we consider self-consistently the large-scale (non-wavelike) equilibrium tidal flow in a homogeneous body.
We assume that the companion is a point mass, moving on an aligned circular orbit around the star with the angular velocity $\Omega_{orb} \boldsymbol{1}_z$ (as depicted in Fig. \ref{fig:sketch}).   
Thus, the dominant component of the tidal potential has the spherical harmonic degree $l=2$ and azimuthal order $m=2$ \citep{ogilvie2014tidal}. 
In the frame rotating with the fluid at the rate $\Omega_s$, the resulting (dimensionless) flow is in the $xy$-plane and takes the form \citep[e.g.][]{barker2013non}
\begin{equation}
\boldsymbol{U}_0 = -\frac{\omega_t \beta}{2}
	\begin{pmatrix}
	\sin (\omega_t t) & \cos (\omega_t t) \\
    \cos(\omega_t t) & -\sin (\omega_t t) \\
	\end{pmatrix}
	\begin{pmatrix}
	x \\
    y \\
	\end{pmatrix},
    \label{eq:U0}
\end{equation}
where $\beta \ll 1$ is the dimensionless tidal amplitude (roughly the ratio of tidal displacement to unperturbed radius), $\omega_t = 2 \, (E^{-1} - E_{orb}^{-1})$ is the dimensionless forcing frequency and $E_{orb}^{-1} = (\Omega_{orb} R^2)/\nu$ is the dimensionless orbital frequency.

\subsection{Numerical modeling}
We follow the numerical implementation introduced in \citet{vidal2020turbulent} to account for tidal flows.  
The non-linear equations (\ref{eqModeling:GeneralEquations}) are solved in their weak variational form by using the spectral-element code Nek5000 \citep[e.g.][]{fischer2007simulation}. 
The computational domain is decomposed into $3584$ non-overlapping hexahedral elements. 
Within each element, the velocity (and pressure) is represented as Lagrange polynomials of order $\mathcal{N}$ (respectively, $\mathcal{N}-2$) on the Gauss-Lobatto-Legendre (Gauss-Legendre) points. 
Temporal discretization is accomplished by a third-order method, based on an adaptive and semi-implicit scheme in which the non-linear and Coriolis terms are treated explicitly, and the remaining linear terms are treated implicitly. 
Solutions are de-aliased following the $3/2$ rule, such that $3\mathcal{N}/2$ grid points are used in each dimension for the non-linear terms, whereas only $\mathcal{N}$ points are used for the linear terms.
We have checked the numerical accuracy in targeted simulations by varying the polynomial order from $\mathcal{N}=7$ to $\mathcal{N}=9$. 

The efficiency of tidal dissipation is investigated by
computing an effective volume-averaged viscosity coefficient $\nu_E$,  introducing the volume average $\langle \cdot \rangle_V = (1/V) \int_V \cdot \, \mathrm{d} V$. 
The forcing amplitude $\beta$ must be large enough to obtain a measurable tidal response, but too large values could strongly modify the results when the amplitude of the tidal flow is much larger than the convective flow \citep[e.g. see in][]{penev2009direct,duguid2020tidal}. 
Only small differences in the properties of the convection have been found for the values of $\beta$ considered below (always smaller than a few percent for the volume-averaged quantities when $\beta \leq 5\times 10^{-2}$, not shown).

Finally, we initiated the convection with random noise to the temperature field and let it saturate without tides (i.e. $\beta=0$) for most of the simulations, before switching on the equilibrium tidal flow. 
We have checked that initiating the convection together with the tidal flow does not lead to noticeably different results. 

\section{Unperturbed convection}
\label{sec:conv}
\begin{table}
    \centering
    \caption{Characteristics of (unperturbed) DNS with NS conditions. Rayleigh number $Ra$, Prandtl number $Pr$, convective velocity $u_{cv}$, and turbulent length scale $l_E$.}
    \begin{tabular}{lccc}
        \hline
        \footnotesize
		$Ra, Pr$ & $E$ & $u_{cv}$ & $l_E$ \\[1ex]
		\hline
        $1 \times 10^5, 1.0$ & $+\infty$ & $(1.84 \pm 0.1) \times 10^1$ & $(4.6 \pm 0.8 ) \times 10^{-1}$ \\
        $3 \times 10^5, 1.0$ & $+\infty$ & $(3.18 \pm 0.2) \times 10^1$ & $(3.9 \pm 1.1 ) \times 10^{-1}$ \\
        $6 \times 10^5, 1.0$ & $+\infty$ & $(4.40 \pm 0.2) \times 10^1$ & $(3.5 \pm 1.1 ) \times 10^{-1}$ \\
        $1 \times 10^6, 1.0$ & $+\infty$ & $(5.64 \pm 0.3) \times 10^1$ & $(3.2 \pm 1.0 ) \times 10^{-1}$ \\
        $2 \times 10^6, 1.0$ & $+\infty$ & $(7.74 \pm 0.4) \times 10^1$ & $(2.9 \pm 1.0 ) \times 10^{-1}$ \\
        $4 \times 10^6, 1.0$ & $+\infty$ & $(1.00 \pm 0.1) \times 10^2$ & $(2.5 \pm 0.8 ) \times 10^{-1}$ \\
        $8 \times 10^6, 1.0$ & $+\infty$ & $(1.31 \pm 0.1) \times 10^2$ & $(2.2 \pm 0.2 ) \times 10^{-1}$ \\
        $1 \times 10^6, 0.3$ & $+\infty$ & $(1.43 \pm 0.1) \times 10^2$ & $(2.5 \pm 0.9 ) \times 10^{-1}$ \\
        $1 \times 10^6, 1.0$ & $10^{-1}$ & $(5.64 \pm 0.3) \times 10^1$ & $(3.1 \pm 1.0 ) \times 10^{-1}$  \\
        $1 \times 10^6, 1.0$ & $10^{-2}$ & $(4.86 \pm 0.3) \times 10^1$ & $(2.9 \pm 1.0 ) \times 10^{-1}$ \\
        \hline
    \end{tabular}
    \label{tab:DNS}
\end{table}

\begin{figure}
    \centering
    \includegraphics[width=0.49\textwidth]{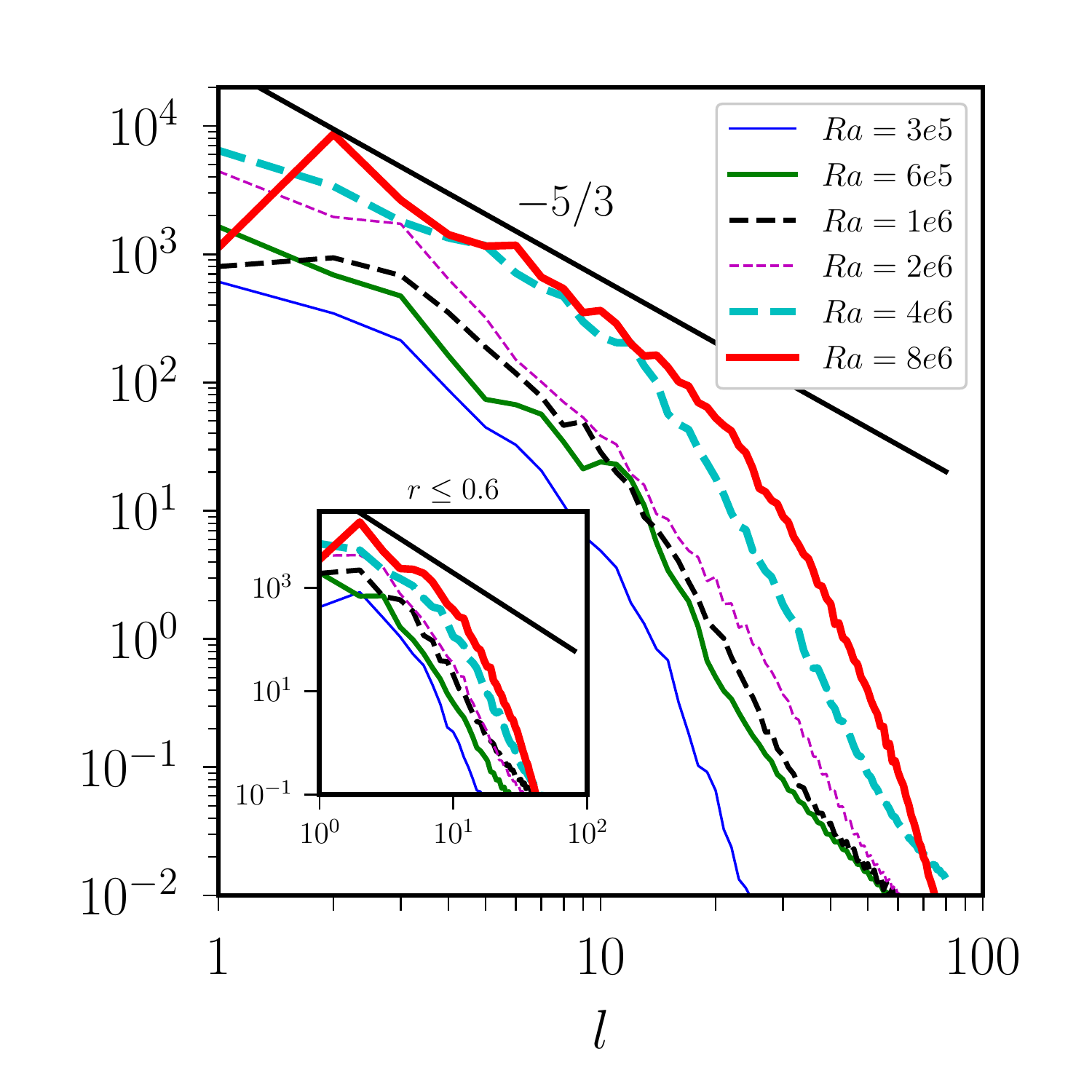}
    \caption{Instantaneous volume-averaged spectrum of the kinetic energy, as a function of the spherical harmonic degree $l\geq 1$ (using orthonormalized spherical harmonics). Thick black line shows the Kolmogorov power law $l^{-5/3}$. 
    Spectra have been computed by interpolating the data to a spherical grid, and then by performing a spherical harmonics analysis \citep{schaeffer2013efficient}. Inset shows the volume-averaged spectra restricted to $r \leq 0.6$.
   }
    \label{fig:speclunperturbed}
\end{figure}

We simulate highly super-critical convection with $Ra \gg Ra_c$ and $Pr=1$, where the critical value for linear onset, computed using a dedicated linear solver \citep{vidal2015quasi,monville2019rotating}, is $Ra_c = 4019$ with NS conditions \citep[the latter value corrects the onset given in][which corresponds instead with the critical value for stress-free BC]{vidal2020turbulent}.   
The parameters and outputs for the DNS with $\beta=0$ are summarized in table \ref{tab:DNS}. 
The spatial spectrum of the unperturbed convection is illustrated in Fig. \ref{fig:speclunperturbed}.
The spectra are well converged with our adopted resolution and they exhibit (small) inertial-like ranges, with a Kolmogorov scaling ($\propto -5/3$) that emerges more clearly when $Ra$ is increased.

\begin{figure}
    \centering
    \includegraphics[width=0.49\textwidth]{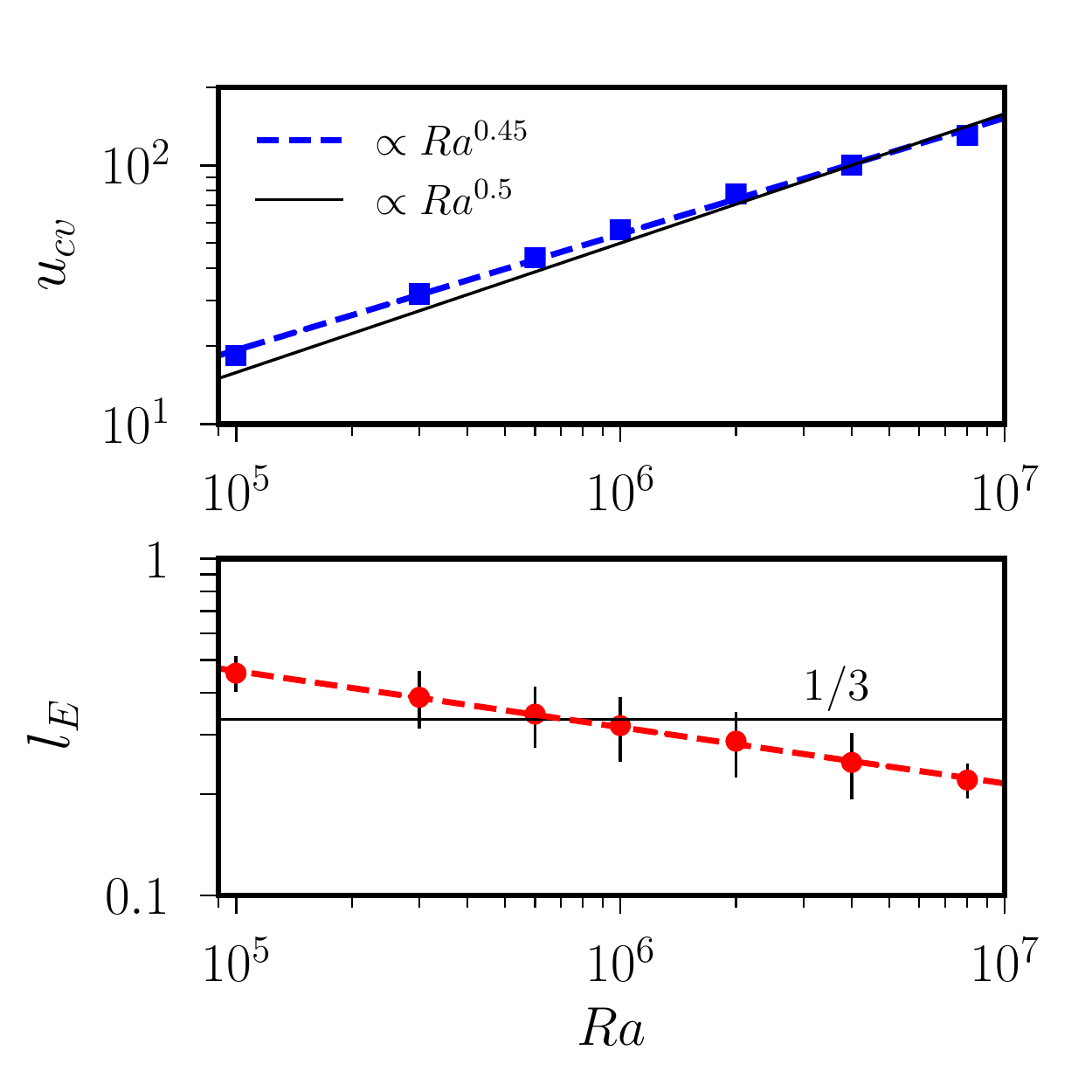} \\
    \caption{\emph{Top panel}: Convective velocity $u_{cv}$ as a function of $Ra$ in DNS. 
    \emph{Bottom panel}: Length scale $l_E$ as a function of $Ra$.
    Dashed line is the power law $l_E = 3.17 \, Ra^{-0.17}$. 
    Horizontal line indicates the value $l_E=1/3$ considered in \citet{vidal2020turbulent} with $Ra=10^6$.}
    \label{fig:figlE}
\end{figure}

For astrophysical applications, the convection is often characterized using MLT by the (unperturbed) turbulent viscosity $\nu_{cv} \sim u_{cv} \, l_E$, with a typical amplitude of the flow $u_{cv}$ and a typical length scale of the turbulent eddies $l_E$. 
To define the convective velocity $u_{cv}$, we use the volume-averaged root-mean-square radial velocity $u_{cv} = (\langle u_r^2 \rangle_V)^{1/2}$ that characterizes the radial mixing. 
We find $u_{cv} \propto Ra^{0.45}$ in the DNS (top panel in Fig. \ref{fig:figlE}), which is in reasonably good agreement with the MLT scaling $\propto Ra^{1/2}$ expected in the fully turbulent regime \citep[e.g.][]{Spiegel1971}. 
This indicates that the convective velocities in our DNS are in an approximately diffusion-free regime, as is expected in stars and planets.

The length scale $l_E$ is usually defined as a function of the local pressure scale height in stellar interiors, but this definition cannot be self-consistently employed in Boussinesq models.
Estimating $l_E$ in global models is difficult \citep[except for rapidly rotating convection, as reported in][]{guervilly2019turbulent}, but a useful characterization of turbulent flows is the Taylor wavenumber $k_T$ \citep[e.g.][]{rieutord2014fluid}
\begin{equation}
    k_T = \sqrt{\langle |\boldsymbol{\nabla} \times\boldsymbol{u}|^2 \rangle_V / \langle |\boldsymbol{u}|^2 \rangle_V},
    \label{eq:taylorK}
\end{equation}
from which we can estimate a turbulent length scale as $l_E=\pi/k_T$ (based on the half wavelength). 
Note that this scale does not represent the energetically-dominant eddies, but a scale intermediate between the ``outer scale" and the dissipation scales, and fairly represents the mean size of the eddies in the turbulent cascade. 
Indeed, in our DNS that do not possess very long inertial ranges, $l_E$ works reasonably well to define the typical size of the turbulent eddies  (which we have verified by visual inspection of the flow).
We show in Fig. \ref{fig:figlE} (bottom panel) the evolution of $l_E$ as a function of $Ra$ in the DNS, and observe that the length scale displays the power law\footnote{A similar scaling for $l_E$ can be obtained by considering that it should scale like the geometric mean $l_E \sim (\eta R)^{0.5}$, with the outer scale $R\sim 1$ and the dissipation scale $\eta\sim R/Re^{3/4}$ \citep{rieutord2014fluid}, where $Re$ is a Reynolds number of the large-scale eddies (assuming $Re \sim Ra^{0.5}$, consistently with Fig. \ref{fig:figlE}). This gives $l_E \sim Ra^{-0.19}R$ in dimensional units.} $l_E \propto Ra^{-0.17}$. 
It also agrees with the value $l_E \simeq 1/3$ at $Ra=10^6$, which was considered in \citet{vidal2020turbulent}.

\begin{figure}
    \centering
    \includegraphics[width=0.49\textwidth]{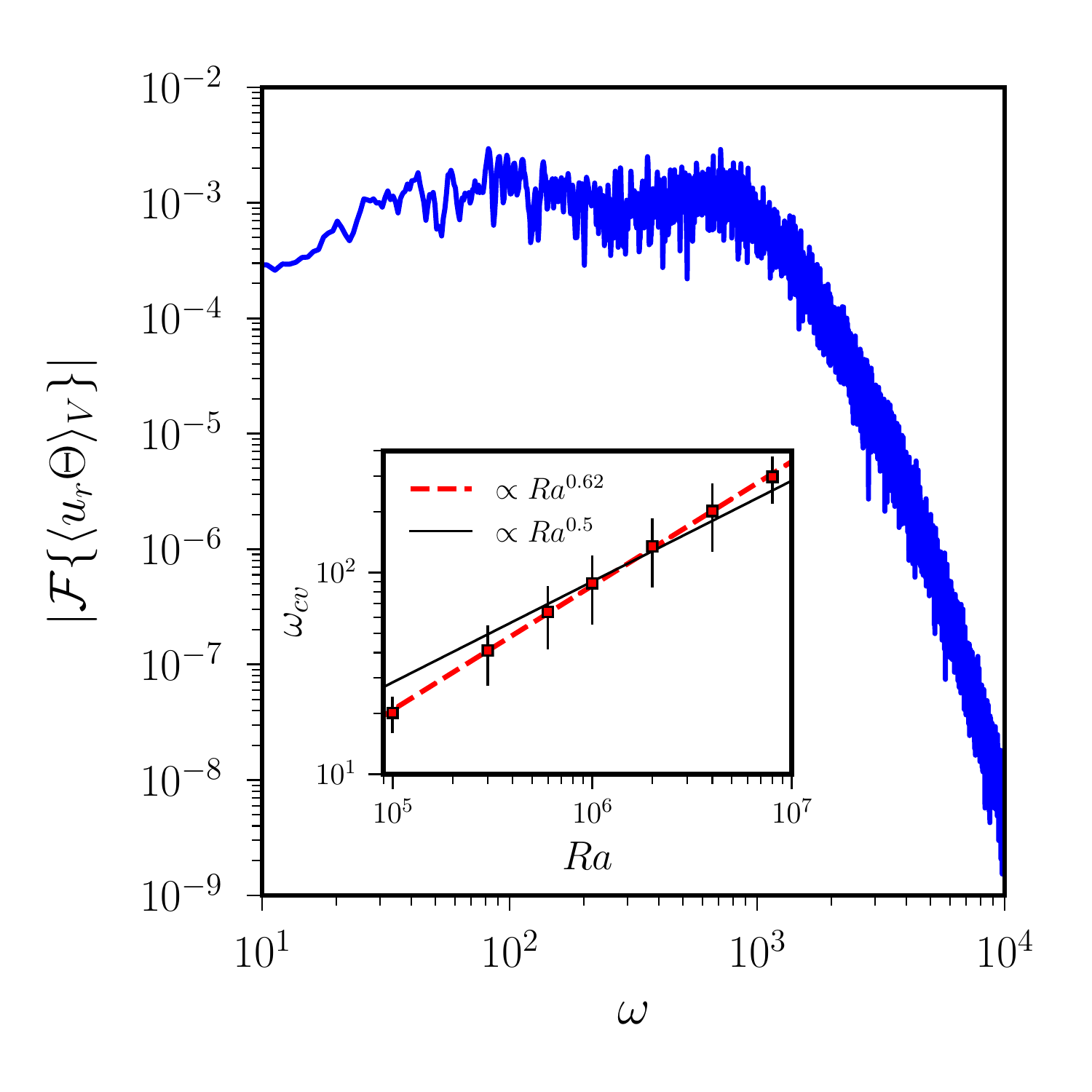}
    \caption{Frequency spectrum of the convective flux $\langle u_r \Theta \rangle_V$ in DNS with $Ra=4 \times 10^6$. 
    Inset panel shows the convective (angular) frequency $\wcv$ as a function of $Ra$.} 
    \label{fig:wcv}
\end{figure}

An estimate of the the convective (angular) frequency $\wcv$ is also required. 
By analogy with stellar models, one can define the convective frequency based on the input parameters \citep[as also considered in][]{ogilvie2012interaction}.  
To do so, we introduce the dimensionless Brunt-V\"ais\"al\"a frequency $N(r)$ given here by $N^2 (r) = -2 r^2 \, Ra/Pr$, and define a typical convective frequency $\omega_{cv} \sim |N_0|$ with the mean radial value $|N_0|= |N^2(1)|^{1/2}/2 \propto (Ra/Pr)^{1/2}$, whose scaling agrees with MLT \citep[e.g.][]{Spiegel1971}. 
Alternatively, a more accurate definition could be based on the turbulent properties of the convective flows.
In the following, we compute the frequency spectrum of the time series $X(t)$ defined as $|\mathcal{F} \{X(t)\}|$, where $\mathcal{F}$ is the Discrete Fourier Transform, as a function of the angular frequency $\omega$. 
We first remove the mean value of the time series and then apply a Hanning window function before we compute numerically the Fourier transform (using the FFT algorithm, and normalizing by the length of the signal). 
One may define $\wcv$ as the frequency that provides the maximum contribution to the convective flux $\langle u_r \Theta \rangle_V$, but the convective frequency is actually poorly constrained from the spectrum of this quantity, which does not exhibit a clearly defined peak (see Fig. \ref{fig:wcv}). 
We choose to instead define the convective frequency as $\wcv = u_E/l_E$, to be consistent with simple MLT expectations.
We find that $\wcv \propto Ra^{0.62}$ in the DNS (see inset), which is quite close to the
MLT prediction $\wcv \propto Ra^{0.5}$ \citep[e.g.][]{Spiegel1971}.

\begin{figure}
    \centering
    \includegraphics[width=0.47\textwidth]{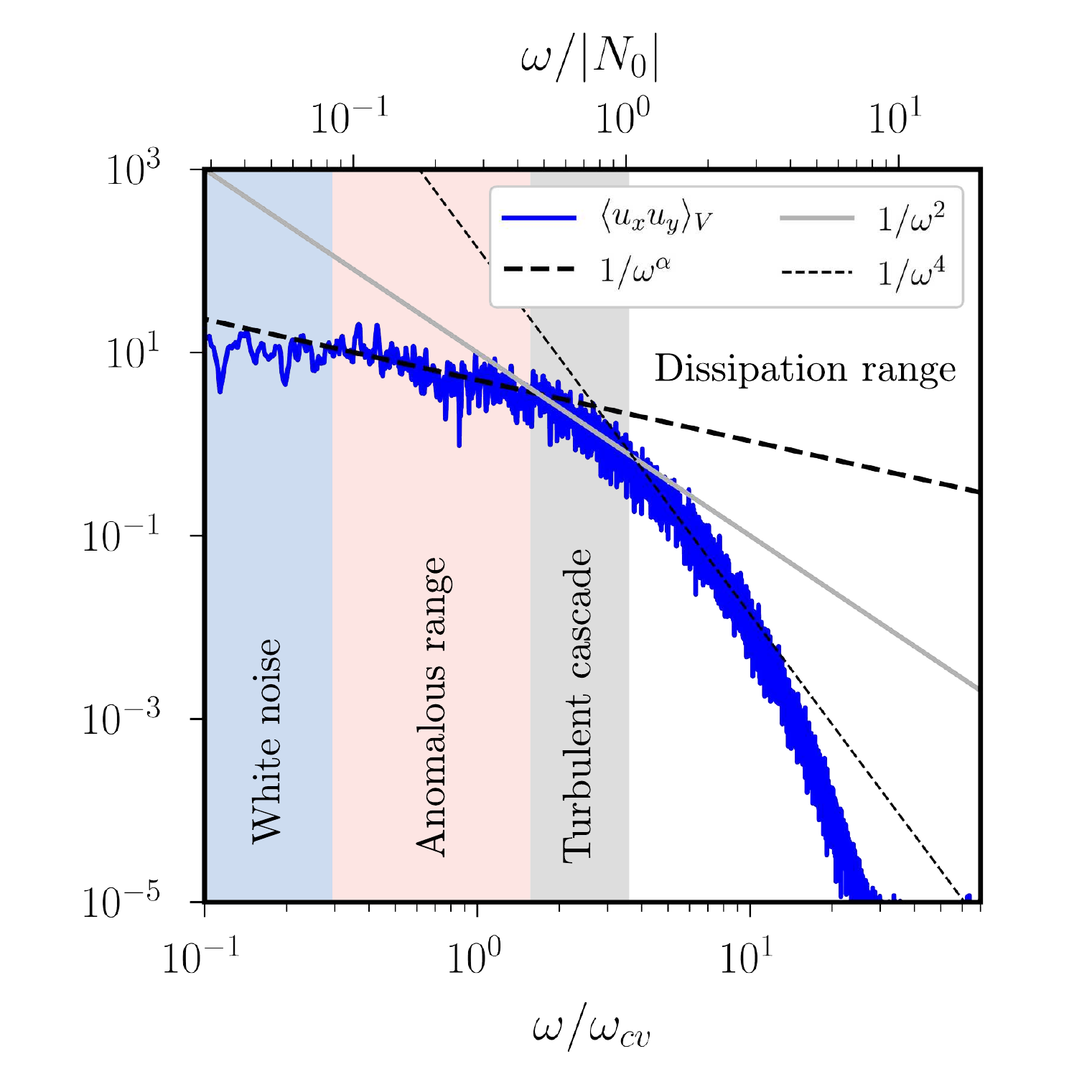}
    \caption{Frequency spectrum of $\langle u_x u_y \rangle_V$ for DNS with $Ra=4 \times 10^6$. The thick dashed line shows the power law $1/\omega^{0.66}$, and the thick gray line indicates the scaling $1/\omega^2$ expected for a Kolmogorov cascade. 
    }
    \label{fig:specFFT}
\end{figure}

We show in Fig. \ref{fig:specFFT} the frequency spectrum of the Reynolds stress component $\langle u_x u_y \rangle_V$, where the angular frequencies have been normalized by $\wcv$ (bottom axis) and $|N_0|$ (top axis), for the illustrative DNS with $Ra = 4 \times 10^6$. 
Several different regimes are observed (which are also relevant for the spectrum of the kinetic energy, not shown).
For very low frequencies $\wcv \lesssim \mathcal{O}(10^{-1})$, we observe frequency-independent white noise.
Within an intermediate frequency range (here $10^{-1} \lesssim \omega/\wcv \leq \mathcal{O}(1)$), denoted below as the anomalous range, the spectrum is characterized by an anomalous $1/\omega^{\alpha}$ power law with exponents $\alpha < 1$ that vary with $Ra$ and $Pr$ in full spheres (as we will discuss further below). 
For larger frequencies $\omega/\omega_c \geq \mathcal{O}(1)$ in the turbulent cascade, the spectrum first displays the power law $1/\omega^2$ expected for Kolmogorov turbulence \citep{Landau1987Fluid,kumar2018applicability}.
Finally, the frequencies belong to the dissipation range of the convection when $\omega/\omega_c \gg 1$, first with the power-law scaling $1/\omega^{4}$ in a narrow frequency interval \citep[as found in laboratory experiments, see in][]{liot2016simultaneous} and then with a steeper decay. 

\section{Efficiency of tidal dissipation}
\label{sec:results}
\subsection{Effective viscosity coefficient}
\label{subsec:eddy}
We primarily extract the turbulent viscosity from our DNS by defining an effective viscosity coefficient $\nu_E$, which is computed by balancing the mean rate at which convection does work on the tidal flow with the mean rate of viscous dissipation of the latter flow \citep[e.g.][]{goodman1997fast,duguid2020tidal,vidal2020turbulent}. 
This leads to $\nu_E = \langle \nu_t (r, \theta, \phi) \rangle_V $ with
\begin{subequations}
\begin{equation}
   \nu_t(r,\theta,\phi) = -\frac{1}{(\omega_t \beta)^2 \Delta T} \int_{t_0}^T  \boldsymbol{u} \boldsymbol{\cdot} \left [ (\boldsymbol{u} \boldsymbol{\cdot} \boldsymbol{\nabla}) \, \boldsymbol{U}_0 \right ] \, \mathrm{d} t
   \label{eq:eddy}
\end{equation}
and the integrand
\begin{multline}
    \boldsymbol{u} \boldsymbol{\cdot} \left [ (\boldsymbol{u} \boldsymbol{\cdot} \boldsymbol{\nabla}) \, \boldsymbol{U}_0 \right ] = -\frac{\omega_t \beta}{2}\left [ \left( u_x^2- u_y^2\right) \sin (\omega_t t) \right . \\
     \left . + 2 \, u_x u_y \cos (\omega_t t) \right], 
\end{multline}
\end{subequations}
where $\Delta T = T-t_0$ is the time-interval used for integration (with $t_0$ being an appropriate initial time in the saturated regime). 
The time average in expression (\ref{eq:eddy}) is obtained by fitting a linear slope to the cumulative time integral to reduce turbulent noise.
\citep[e.g. see fig. 13 in][]{duguid2020tidal}. 
Global simulations in the presence of large-scale tidal flows are very demanding, because they must be run for a sufficiently long duration to reduce noise. We have therefore integrated each simulation with a tidal flow for at least one viscous time unit (i.e.~$\Delta T \geq 1$), corresponding with more than a hundred tidal periods, to obtain converged statistics for the effective viscosity. 
Finally, since the background flow strictly does not satisfy the boundary conditions in a sphere, we have verified that the volume average is not dominated by regions near the boundary, and is instead due to interactions with turbulent flows in the bulk \citep[not shown here, but see fig. 5 in][]{vidal2020turbulent}. 

\begin{figure}
    \centering
    \begin{tabular}{c}
    \includegraphics[width=0.47\textwidth]{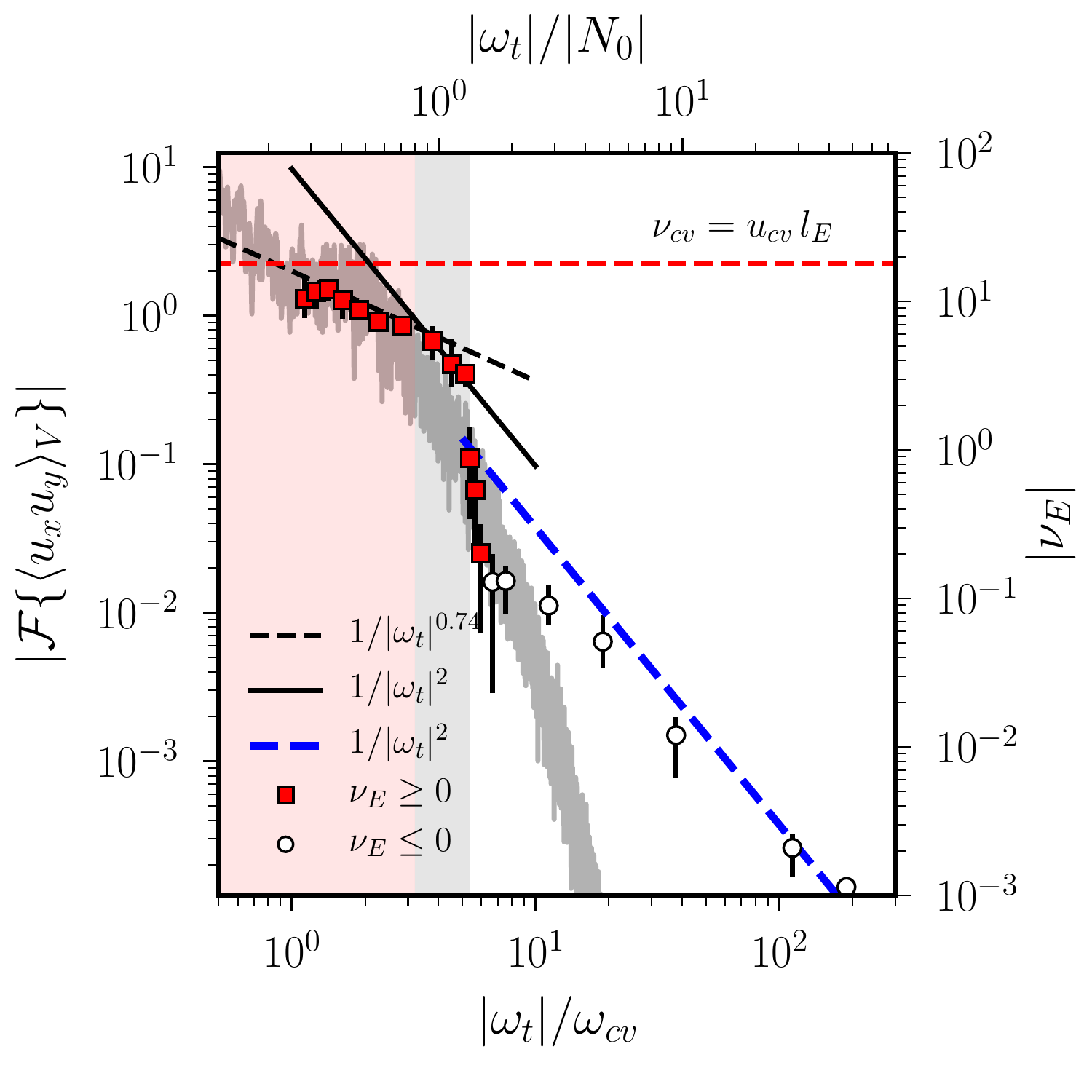} \\
    (a) $Ra=10^6$ \\
    \includegraphics[width=0.47\textwidth]{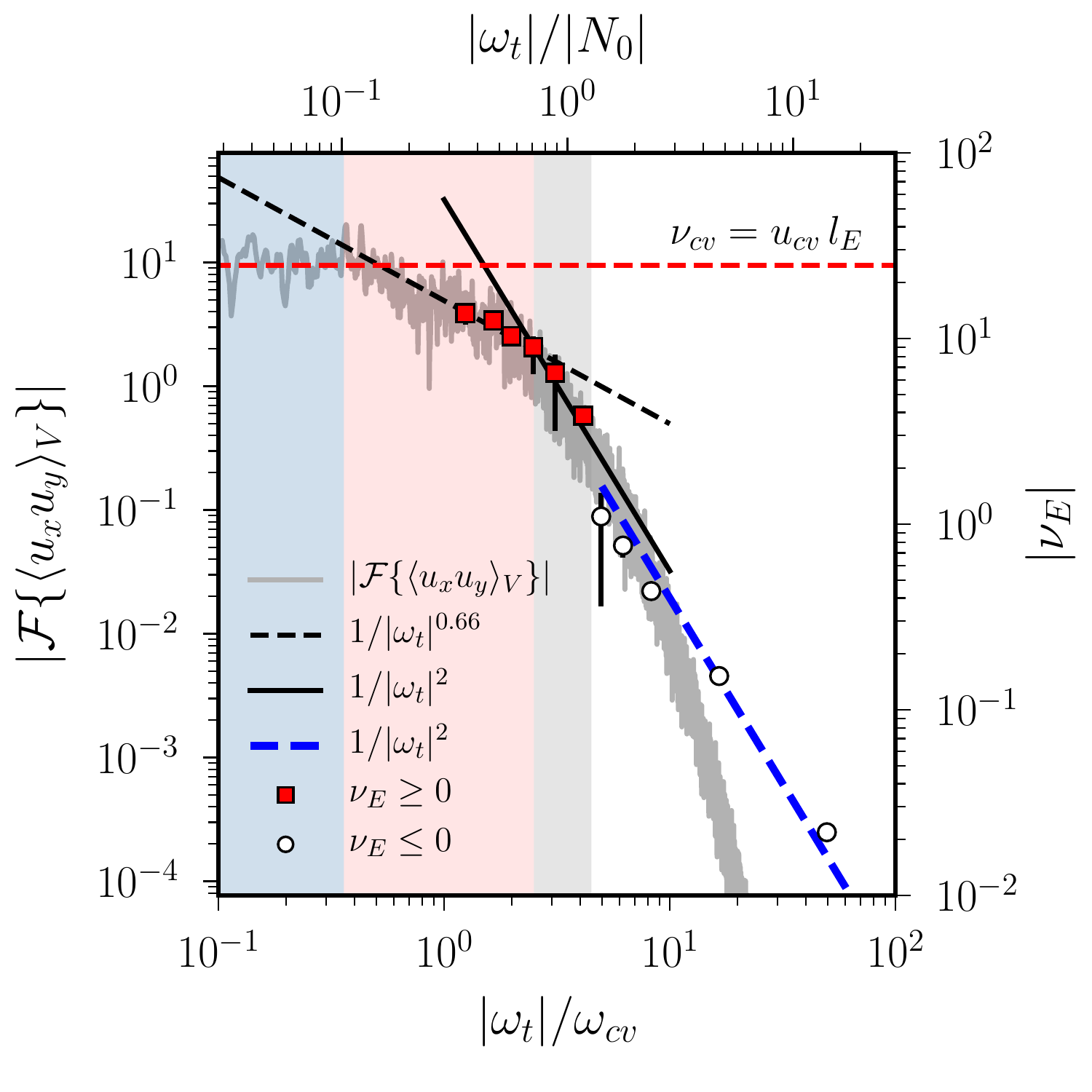} \\
    (b) $Ra=4 \times 10^6$ \\
    \end{tabular}
    \caption{Direct measurements of the effective viscosity $\nu_E$ in non-rotating DNS with $Pr=1$ and $\beta=5 \times 10^{-2}$, as a function of $|\omega_t|/\wcv$ or $|\omega_t|/|N_0|$. Squares: $\nu_E > 0$. Circles: $\nu_E < 0$.
    Horizontal dashed lines: MLT expectation $\nu_{cv} = u_{cv} \, l_E$ in the low-frequency regime (i.e. $|\omega_t|\ll \wcv$). 
    The gray curve shows the frequency spectrum of $\langle u_x u_y \rangle_V$ for unperturbed convection with $\beta=0$, as a function of the scaled angular frequency $|\omega|/\wcv$ (same horizontal values as $|\omega_t|/\wcv$).
    Background colors refer to figure \ref{fig:specFFT}.}
    \label{fig:eddy1}
\end{figure}

We show in Fig. \ref{fig:eddy1} the direct computations of $\nu_E$ in the DNS with $Ra=10^6$ and $Ra=4\times10^6$, assuming a tidal amplitude of $\beta=5 \times 10^{-2}$ (which is e.g. a relevant value for a solar-mass binary in a one-day orbit). 
We also over-plot the frequency spectrum of the Reynolds stress component $\mathcal{F}\left\{\langle u_xu_y\rangle_V\right\}$ as the gray lines in both panels.
The clearest result evident in Fig. \ref{fig:eddy1} is that $\nu_E$  decreases as the ratio $|\omega_t|/\wcv$ is increased, which means that the efficiency of the dissipation is reduced for fast tides.
For the particular DNS with $Ra=10^6$, the two canonical frequency-reduction laws (linear and quadratic) are approximately obtained, which were previously discussed in \citet{vidal2020turbulent}. 
However, our more thorough analysis reveals that the frequency-reduction law follows several successive power laws that are in good agreement with the frequency spectrum of the unperturbed convection. 
Within the anomalous (intermediate-frequency) range where the frequency spectrum of the unperturbed convection varies as $1/\omega^\alpha$ (with power exponents $\alpha < 1$ in full spheres), the viscosity is reduced as $\nu_E \propto 1/(|\omega_t|/\wcv)^{\alpha}$. 
Then, for higher frequencies in the (narrow) turbulent cascade that displays the Kolmogorov power law $1/\omega^{2}$, the previous frequency-reduction scaling ceases to be valid and is replaced by a quadratic reduction for the effective viscosity (with only positive values). 

Therefore, we obtain two successive frequency-reduction laws for the effective viscosity for frequencies below the dissipation range of the turbulence. 
Although our turbulent cascade corresponds here to a narrow frequency interval, our results confirm that a Kolmogorov spectrum is associated with a quadratic reduction \citep[as postulated by][though it is unclear whether their mechanism applies in detail]{goldreich1977turbulent}. 
However, for smaller frequencies in the anomalous range (i.e. outside the turbulent cascade), the frequency reduction of the effective viscosity is neither quadratic nor linear, but follows instead the anomalous frequency spectrum $1/\omega^\alpha$ of the convection.
Moreover, the power exponent $\alpha < 1$ is reduced in full spheres when the Rayleigh number is increased, for instance with $\alpha=0.74 \pm 0.05$ when $Ra=10^6$ and $\alpha=0.66 \pm 0.05$ when $Ra=4 \times 10^6$.

\begin{figure}
    \centering
    \includegraphics[width=0.47\textwidth]{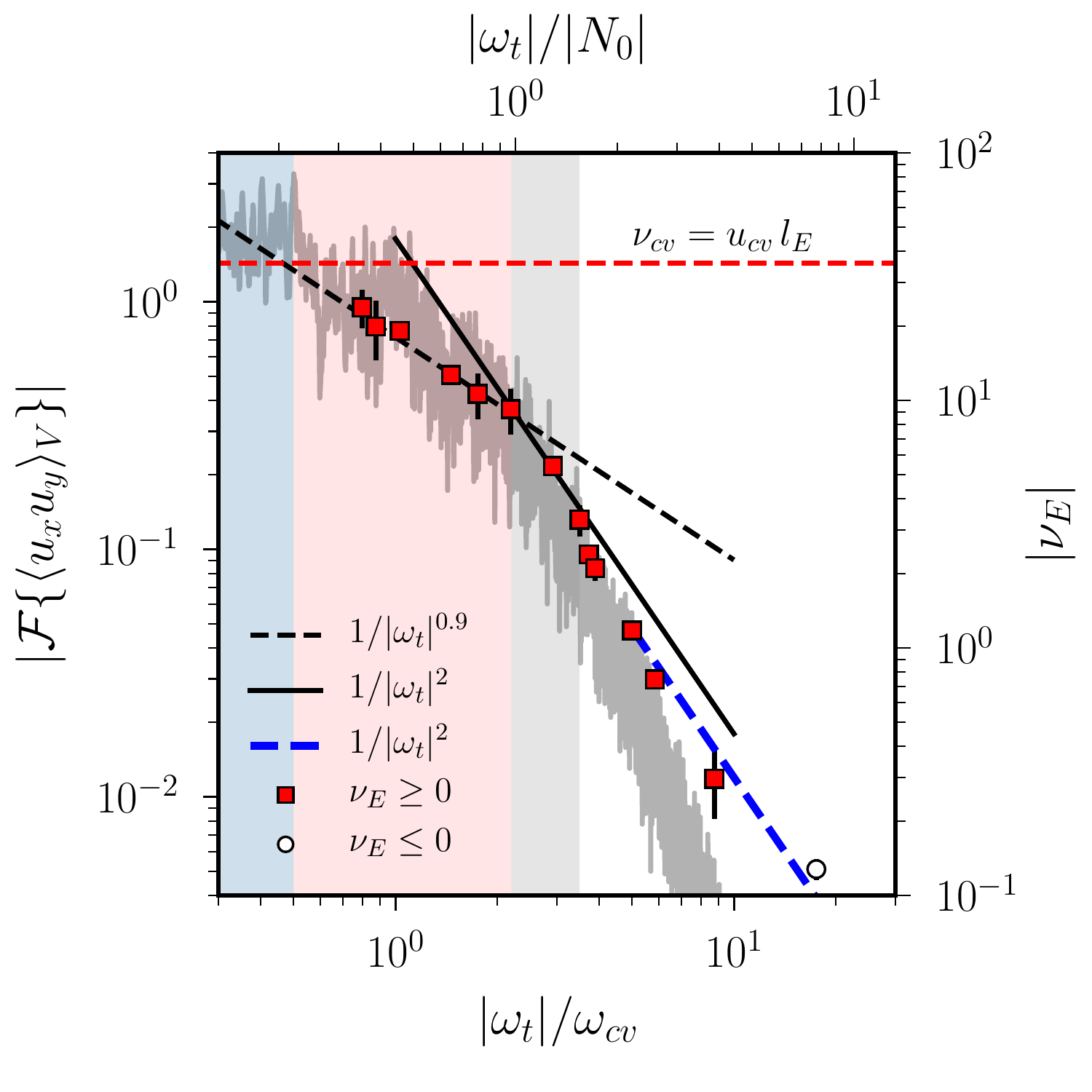}
    \caption{Direct measurements of the effective viscosity $\nu_E$ in non-rotating DNS as a function of $|\omega_t|/\wcv$ (or $|\omega_t|/|N_0|$). DNS with $Ra=10^6$, $Pr=0.3$ and $\beta = 5 \times 10^{-2}$.
    Squares: $\nu_E > 0$. Circles: $\nu_E < 0$. 
    Horizontal dashed lines: MLT prediction $\nu_{cv} = u_{cv} \, l_E$. 
    The gray curve shows the frequency spectrum of $\langle u_x u_y \rangle_V$ for unperturbed convection with $\beta=0$, as a function of the scaled angular frequency $|\omega|/\wcv$ (same horizontal values as $|\omega_t|/\wcv$).
    Background colors refer to Fig. \ref{fig:specFFT}.}
    \label{fig:eddy2}
\end{figure}

We show in Fig. \ref{fig:eddy2} the effective viscosity measured in DNS with $Ra=10^6$ and $Pr=0.3$ \citep[this is relevant for liquid metals, e.g.][]{kaplan2017subcritical}. Exploring cases with smaller $Pr$ is important because $Pr$ in stellar or planetary convection zones is much smaller than unity \citep[e.g.][]{hanasoge2014quest}.
The frequency range of the Kolmogorov cascade is slightly larger in this case compared to Fig. \ref{fig:eddy1}, and more importantly the transition between positive and negative values occurs at larger tidal frequencies (within the dissipation range) but always when $|\nu_E| \lesssim \nu$. 
Note also that the value of the exponent $\alpha$ is different in the intermediate-frequency regime, showing that $\alpha$ also depends on $Pr$, which indicates a parameter-dependence within the anomalous range.

Note that we have been unable to accurately determine $\nu_E$ in the low-frequency regime ($|\omega_t| \lesssim \wcv$) with these simulations. This is because the amplitude of the tidal flow in this regime was too weak to give a sufficiently strong signal-to-noise ratio.
A crude extrapolation of our results into the low frequency regime is broadly consistent with expectations from MLT though, which would predict $\nu_E \propto \nu_{cv} \sim u_{cv} \, l_E$ when $|\omega_t| \to 0$. The proportionality constant is often assumed to be 1/3 without rigorous justification \citep[e.g][]{zahn1989tidal,OL2007}, based on the analogy with kinetic theory for a microscopic viscosity. 
Here we instead find values close to 1, or in fact in excess of 1 if $\nu_E$ continues to follow the spectrum for smaller $|\omega_t|/\wcv$, indicating more efficient dissipation at low frequencies from this mechanism than the naive application of MLT would predict. This result is broadly consistent with local simulations \citep{duguid2020tidal}, and prior theoretical work obtained with an idealized turbulence model \citep[e.g.][]{goldman2008effective}.  

\subsection{Negative values}
\begin{figure}
    \centering
    \includegraphics[width=0.49\textwidth]{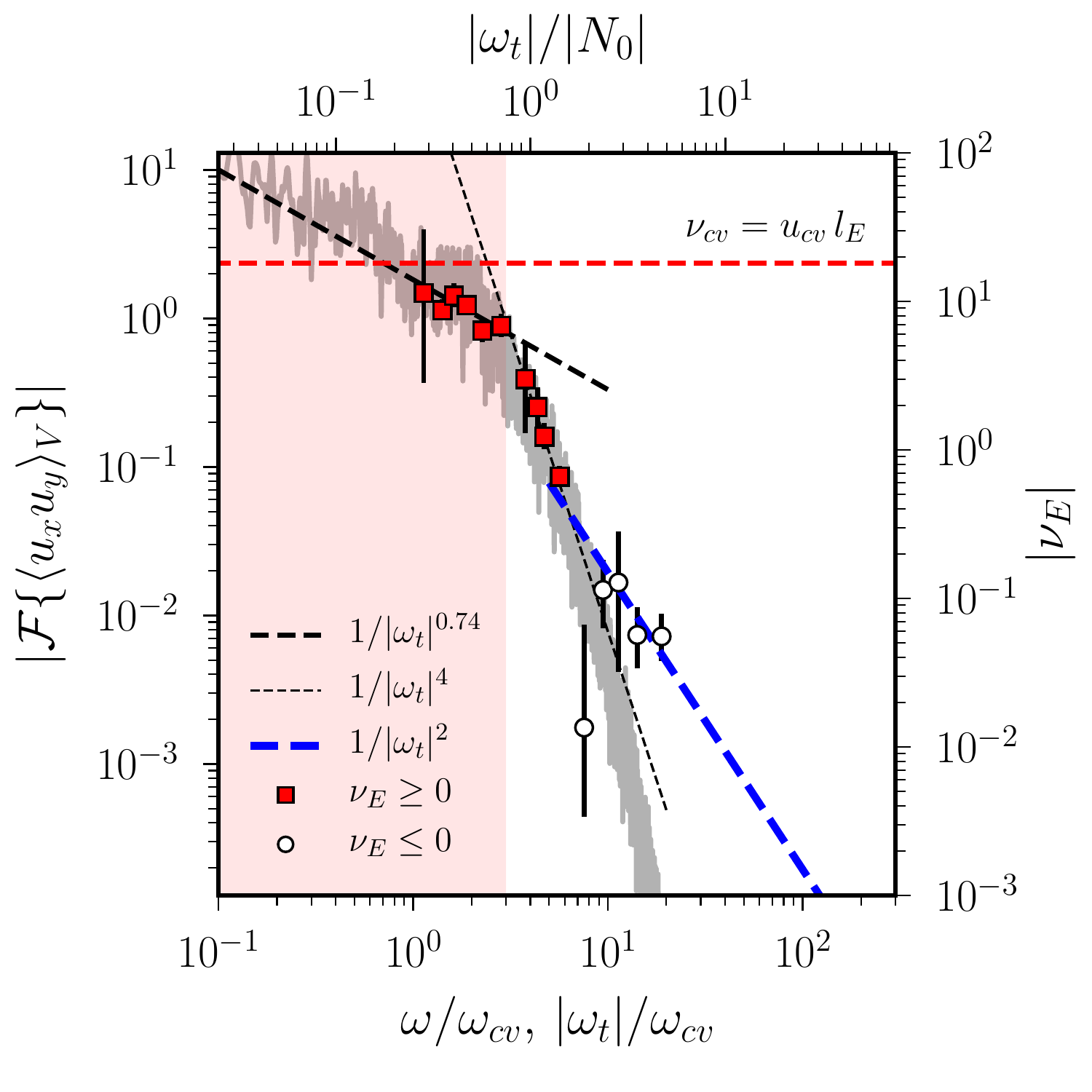}
    \caption{Direct measurements of the effective viscosity $\nu_E$ in non-rotating DNS as a function of $|\omega_t|/\wcv$ (or $|\omega_t|/|N_0|$), showing the transition to negative values of $\nu_E$. DNS with $Ra=10^6$, $Pr=1$ and $|\omega_t| \, \beta = 20$. Squares: $\nu_E > 0$. Circles: $\nu_E < 0$. 
    Horizontal dashed lines: MLT prediction $\nu_{cv} = u_{cv} \, l_E$. 
    The gray curve shows the frequency spectrum of $\langle u_x u_y \rangle_V$ for unperturbed convection with $\beta=0$, as a function of the scaled angular frequency $|\omega|/\wcv$ (same horizontal axis as $|\omega_t|/\wcv$). 
    Background colors refer to Fig. \ref{fig:specFFT}.}
    \label{fig:eddy3}
\end{figure}

Statistically significant negative values of the turbulent viscosity are found in Figs \ref{fig:eddy1} and \ref{fig:eddy2} for much higher frequencies within the dissipation range, which are consistent with previous local results and asymptotic theory \citep{ogilvie2012interaction,duguid2020tidal}.  
The transition towards negative values is better illustrated in Fig. \ref{fig:eddy3} using DNS with $Ra=10^6$ and $Pr=1$, but with the fixed amplitude $|\omega_t| \, \beta = 20$ for the tidal flow (instead of fixing $\beta$). 
This allows us to investigate more efficiently the transition between positive and negative values, without disturbing (to the same extent) the frequency spectrum of the convection contrary to Figs \ref{fig:eddy1}-\ref{fig:eddy2} \citep[for which the amplitude of the tidal flow increases when $|\omega_t|$ increases, see fig. 3 in][]{vidal2020turbulent}.
In the narrow frequency interval where the frequency spectrum displays a $1/\omega^4$ power law, we find that the eddy viscosity is reduced by the same amount but has positive values. 
For larger frequencies within the dissipation range, the effective viscosity changes sign and then follows a generic quadratic reduction once $|\nu_E|\lesssim \nu$.

For very high frequencies, our results indicate that $|\nu_E|\propto (|\omega_t|/\wcv)^{-2}$ even for these negative values, consistently with asymptotic theory \citep{ogilvie2012interaction,duguid2020tidal}. 
Moreover, since the change of sign of $\nu_E$ seems to occur when $|\nu_E|/\nu \lesssim 1$ in dimensional units (corresponding with frequencies $|\omega_t|$ firmly within the dissipation range), this
probably indicates that the observed negative values are not astrophysically relevant but result from (necessarily) adopting simulation parameters that are far removed from their astrophysical values (see below).

\subsection{Complementary analysis}
\label{subsec:extended}
\begin{figure}
    \centering
    \begin{tabular}{c}
    \includegraphics[width=0.47\textwidth]{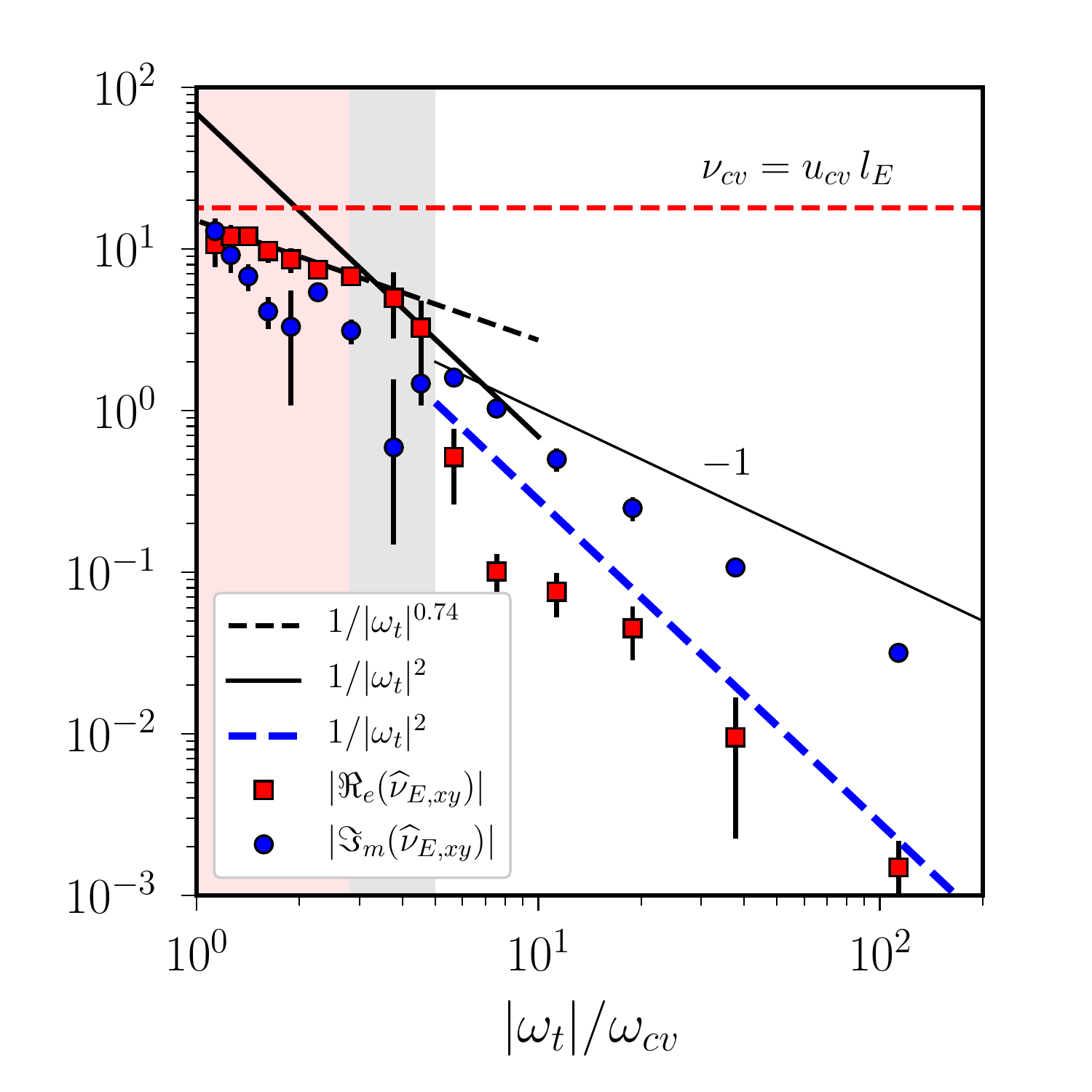} \\
    (a) $Ra=10^6$ \\
    \includegraphics[width=0.47\textwidth]{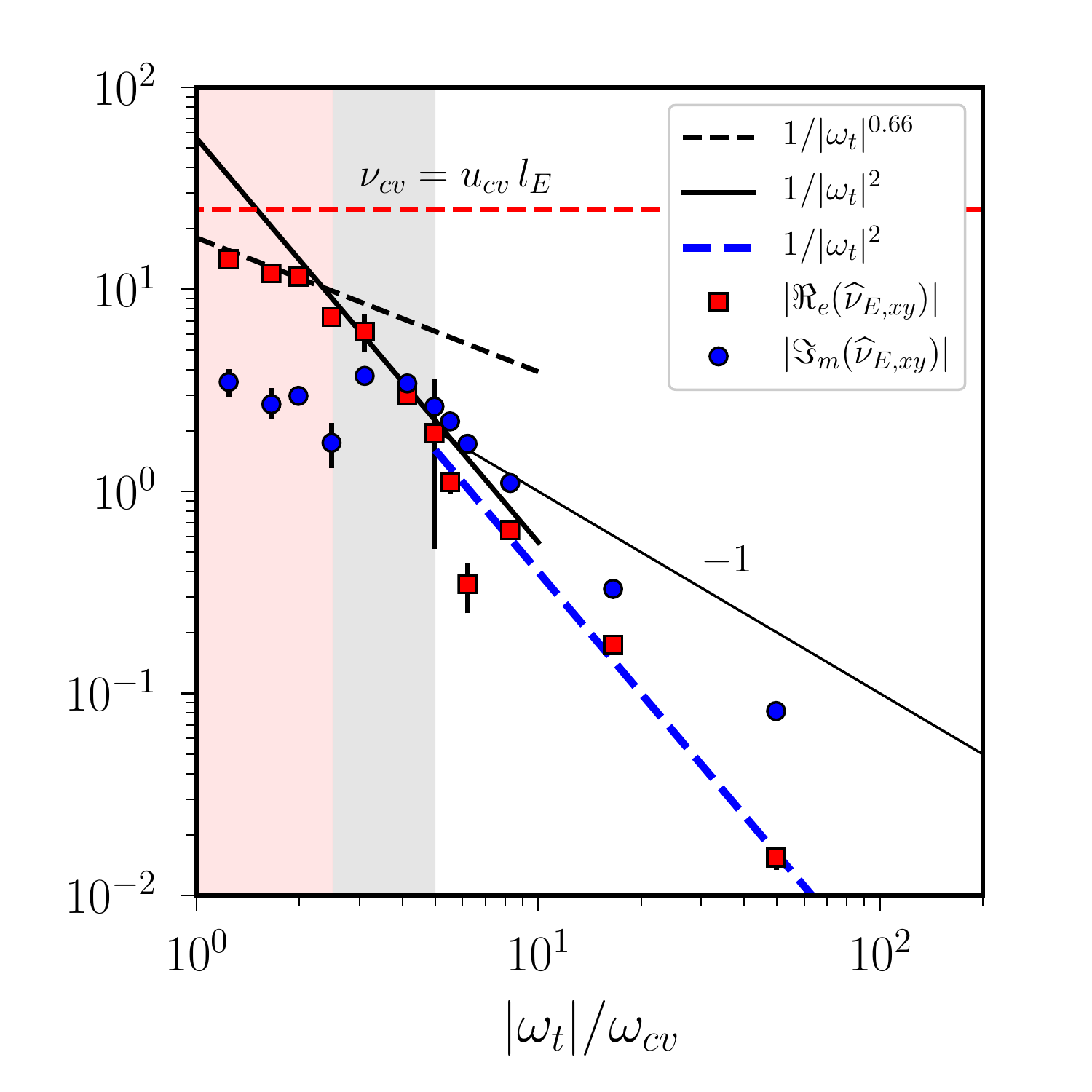} \\
    (b) $Ra=4 \times 10^6$ \\
    \end{tabular}
    \caption{$|\Re_e (\widehat{\nu}_{E,xy})|$ (red squares) and $|\Im_m (\widehat{\nu}_{E,xy})|$ (blue circles) of the contribution to the effective viscosity defined by equation (\ref{eq:closurexy}), as a function of $|\omega_t|/\wcv$ in DNS with $Pr=1$ and $\beta=5 \times 10^{-2}$. 
    Background colors refer to Fig. \ref{fig:specFFT}.}
    \label{fig:closure1}
\end{figure}

We can alternatively compute the effective viscosity associated with each component of the volume-averaged Reynolds stress by relating the stress to the time history of the rate of strain \citep[thus accounting for the oscillatory nature of the tidal flow, see e.g.][]{ogilvie2012interaction}. In the Fourier domain, this gives
\begin{subequations}
\label{eq:closure}
\begin{align}
    \mathcal{F} \{\langle u_x u_y \rangle_V \} &= \widehat{\nu}_{E,xy} \, \omega_t \beta \, \mathcal{F} \{\cos(\omega_t t) \}, \label{eq:closurexy} \\
     \mathcal{F} \{\langle u_x^2 \rangle_V \} &= \widehat{\nu}_{E,xx} \, \omega_t \beta \, \mathcal{F} \{\sin(\omega_t t) \}, \label{eq:closurexx}
\end{align}
\end{subequations}
and similarly for $\mathcal{F} \{\langle u_y^2 \rangle_V \}$, where $[\widehat{\nu}_{E,xy}, \widehat{\nu}_{E,xx}]$ are complex-valued quantities.

In the regime of high-frequency tidal forcing ($|\omega_t| \gg \wcv$), \citet{ogilvie2012interaction} and \citet{duguid2020tidal} used asymptotic theory to demonstrate the visco-elastic nature of the tidal response (using a simple oscillatory shear) for quantity (\ref{eq:closurexy}).
In the latter expression, the real part $\Re_e(\widehat{\nu}_{E,xy})$ represents a turbulent viscosity (which is in phase with the tidal shear and out of phase with the tidal displacement) associated with this component of the flow, which provides a contribution to the total $\nu_E$. 
Asymptotic theory indicates that for high frequencies, this quantity should scale as $|\omega_t|^{-2}$ (with possibly negative values). 
On the other hand, the imaginary part $\Im_m(\widehat{\nu}_{E,xy})$ is related to an effective elasticity (which is out of phase with the tidal shear and in phase with the tidal displacement) and should obey a linear reduction $|\omega_t|^{-1}$ in that regime (indicating an effective elastic modulus that is independent of frequency). 
We show in Fig. \ref{fig:closure1} direct computations of $\widehat{\nu}_{E,xy}$ from equation (\ref{eq:closurexy}) for two different values of $Ra$, which confirm the universal nature of the visco-elastic response of $\langle u_x u_y \rangle_V$ at high tidal frequencies (here $|\omega_t|/\wcv \geq \mathcal{O}(10)$). 
We broadly obtain a linear reduction $|\Im_m(\widehat{\nu}_{E,xy})| \propto |\omega_t|^{-1}$ in the high-frequency regime, and we also recover the expected scaling in $|\omega_t|^{-2}$ for the turbulent viscosity $|\Re_e(\widehat{\nu}_{E,xy})|$ in this regime. 
The latter is always smaller than $|\Im_m(\widehat{\nu}_{E,xy})|$, indicating a primarily elastic response to high frequency shear, with a weaker viscous component.

However, this asymptotic theory does not apply for the lower forcing frequencies $|\omega_t|/\wcv \leq 10$ that we consider here. Indeed, for these lower frequencies, $|\Re_e(\widehat{\nu}_{E,xy})|$ and $|\Im_m(\widehat{\nu}_{E,xy})|$ have comparable magnitudes, and the viscous component can even dominate. Hence, the predictions of the asymptotic theory cannot be strictly invoked to support the quadratic reduction for lower frequencies than those contained in the dissipation range of the convection in our simulations. Instead, we find that the $\nu_E$ behaves similarly to the frequency spectrum of $\langle u_x u_y \rangle _V$ (e.g. Fig. \ref{fig:eddy1}), indicating that this is a key quantity governing the frequency-reduction of the eddy viscosity in our simulations. 

\begin{figure}
    \centering
    \includegraphics[width=0.47\textwidth]{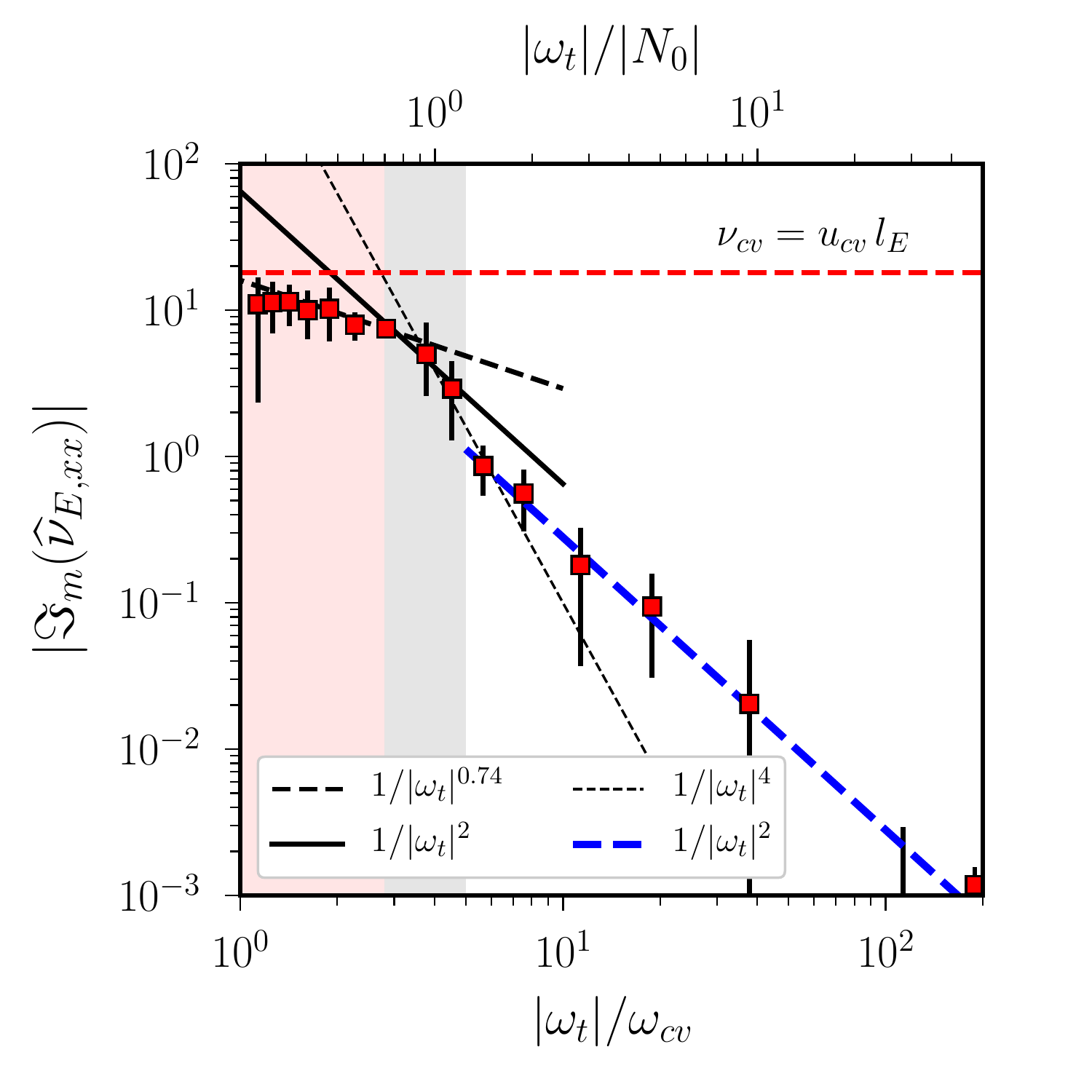}
    \caption{Effective viscosity contribution $|\Im_m (\widehat{\nu}_{E,xx})|$ computed from expression (\ref{eq:closurexx}), as a function of $|\omega_t|/\wcv$ in DNS with $Ra=10^6$, $Pr=1$ and $\beta=5 \times 10^{-2}$. 
    Background colors refer to Fig. \ref{fig:specFFT}.}
    \label{fig:closure2}
\end{figure}

We also illustrate in Fig. \ref{fig:closure2} the contribution to $\nu_E$ from $\Im_m (\widehat{\nu}_{E,xx})$ computed from (\ref{eq:closurexx}). 
Similar results are obtained for the $\langle u_y^2\rangle_V$ component (since $u_x$ and $u_y$ play symmetrical roles, not shown). 
The amplitude of the effective viscosity contribution from this component is in broad quantitative agreement with Figs \ref{fig:eddy1} and \ref{fig:closure1}, which cross-validates our computations for the turbulent viscosity. This also agrees with \citet{penev2009direct}, who showed that the effects of convective turbulence on a large-scale oscillatory shear flow is fairly well represented by an effective viscosity coefficient.

\subsection{Inclusion of weak rotation}
\label{subsec:rot}
\begin{figure}
    \centering
    \includegraphics[width=0.47\textwidth]{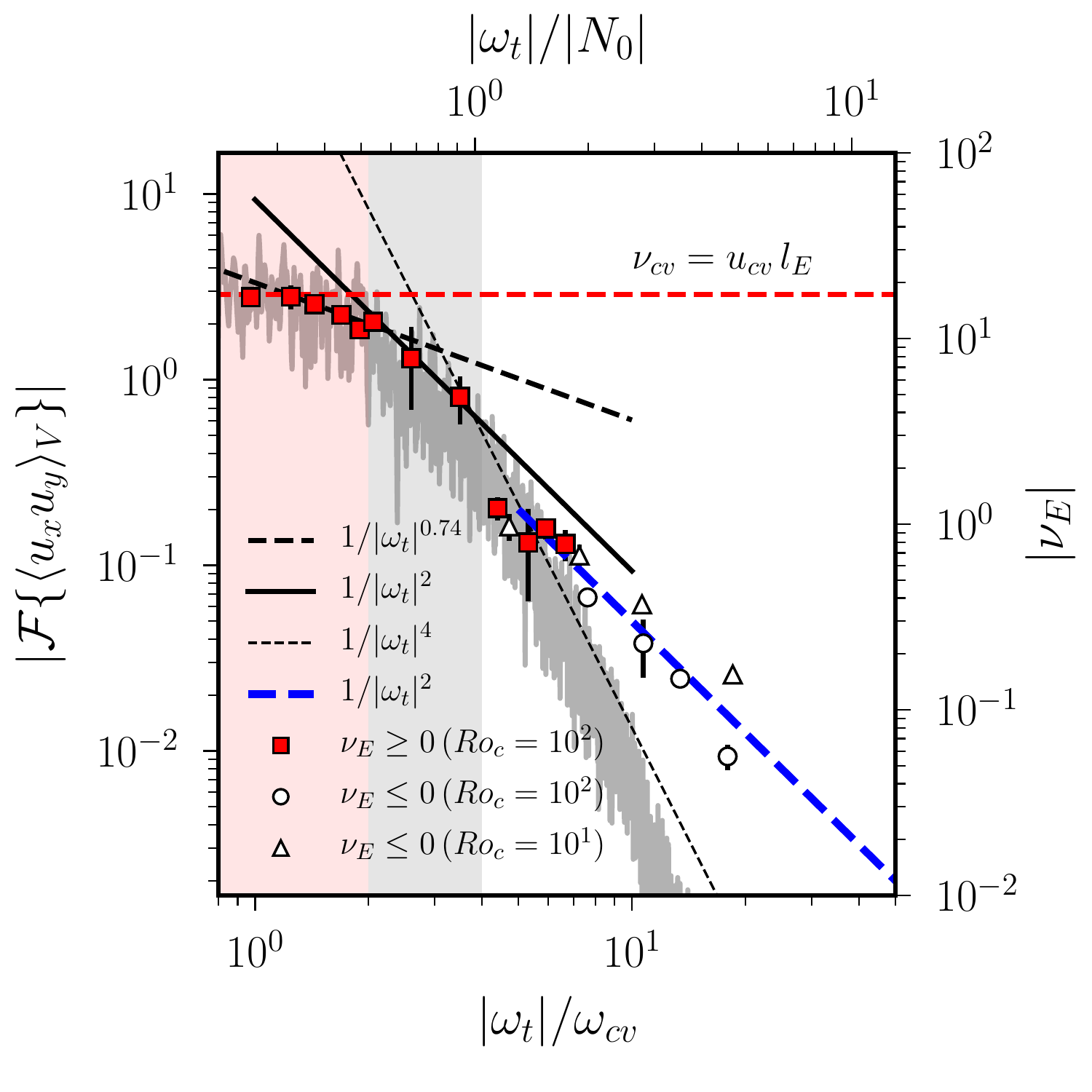}
    \caption{Direct measurements of the effective viscosity $\nu_E$ in weakly rotating DNS with $Ra=10^6$, $Pr=1$ for tidal amplitude $\beta = 5 \times 10^{-2}$, as a function of $|\omega_t|/\wcv$ (or $|\omega_t|/|N_0|$). Red squares: $\nu_E > 0$ in DNS with $Ro_c=10^2$ ($E=10^{-1}$) and $\beta=5 \times 10^{-2}$. Empty circles: $\nu_E < 0$ in DNS with $Ro_c=10^2$ ($E=10^{-1}$) and $\beta=5 \times 10^{-2}$. 
    Empty triangles: $\nu_E < 0$ for DNS with $Ro_c=10^1$ ($E=10^{-2}$) and $\beta=10^{-2}$.
    The horizontal dashed lines indicate the MLT expectation $\nu_{cv} \sim u_{cv} \, l_E$ for $Ro_c=10^2$ ($E=10^{-1}$), and the gray curve shows the frequency spectrum of $\langle u_x u_y \rangle_V$ for unperturbed DNS with $Ro_c=10^2$ ($E=10^{-1}$) as a function of $\omega/\wcv$ (same horizontal axis as $|\omega_t|/\wcv$). 
    Background colors refer to Fig. \ref{fig:specFFT}.}
    \label{fig:eddyRot}
\end{figure}

We now introduce global rotation to assess the robustness of the observed frequency-reduction laws for slowly rotating stars or planets. 
One measure for the degree of rotational constraint in convection-driven flows is given by the convective Rossby number $Ro_c = E \sqrt{Ra/Pr}$. 
Weakly rotating convection is believed to approach non-rotating convection \citep[e.g.][]{gastine2016scaling,long2020scaling}, and so quantitatively similar results are expected for the turbulent viscosity when $Ro_c \gg 1$ (as considered below). 
We show in Fig. \ref{fig:eddyRot} the DNS with $Ro_c = 10^2$ ($E=10^{-1}$) and $Ro_c = 10^1$ ($E=10^{-2}$). 
By comparison with Fig. \ref{fig:eddy1}a, we observe values of $\nu_E$ that are close to the ones obtained in the non-rotating DNS. 

Our results indicate here that weak global rotation does not significantly modify the frequency-reduction laws of $\nu_E$ found in non-rotating spherical convection. 
Yet, rapid rotation is known to strongly affect spherical convection \citep[e.g.][]{guervilly2019turbulent}, and is therefore believed to strongly modify the effective viscosity when $Ro_c\ll 1$ \citep{mathis2016impact}.
Another complication with incorporating rapid rotation in our model is that the tidal (elliptical) instability can be triggered for large enough $\beta$ when $-1 \leq \Omega_{orb}/\Omega_s = E/E_{orb} \leq 3$ \citep{barker2016non,vidal2017inviscid}.
Further work is required to explore this regime, which might be relevant for giant planets or young rapidly rotating stars. 

\section{Discussion}
\label{sec:discussion}
\subsection{Non-Kolmogorov turbulent spectrum}
\begin{figure*}
    \centering
    \begin{tabular}{cc}
    \includegraphics[width=0.38\textwidth]{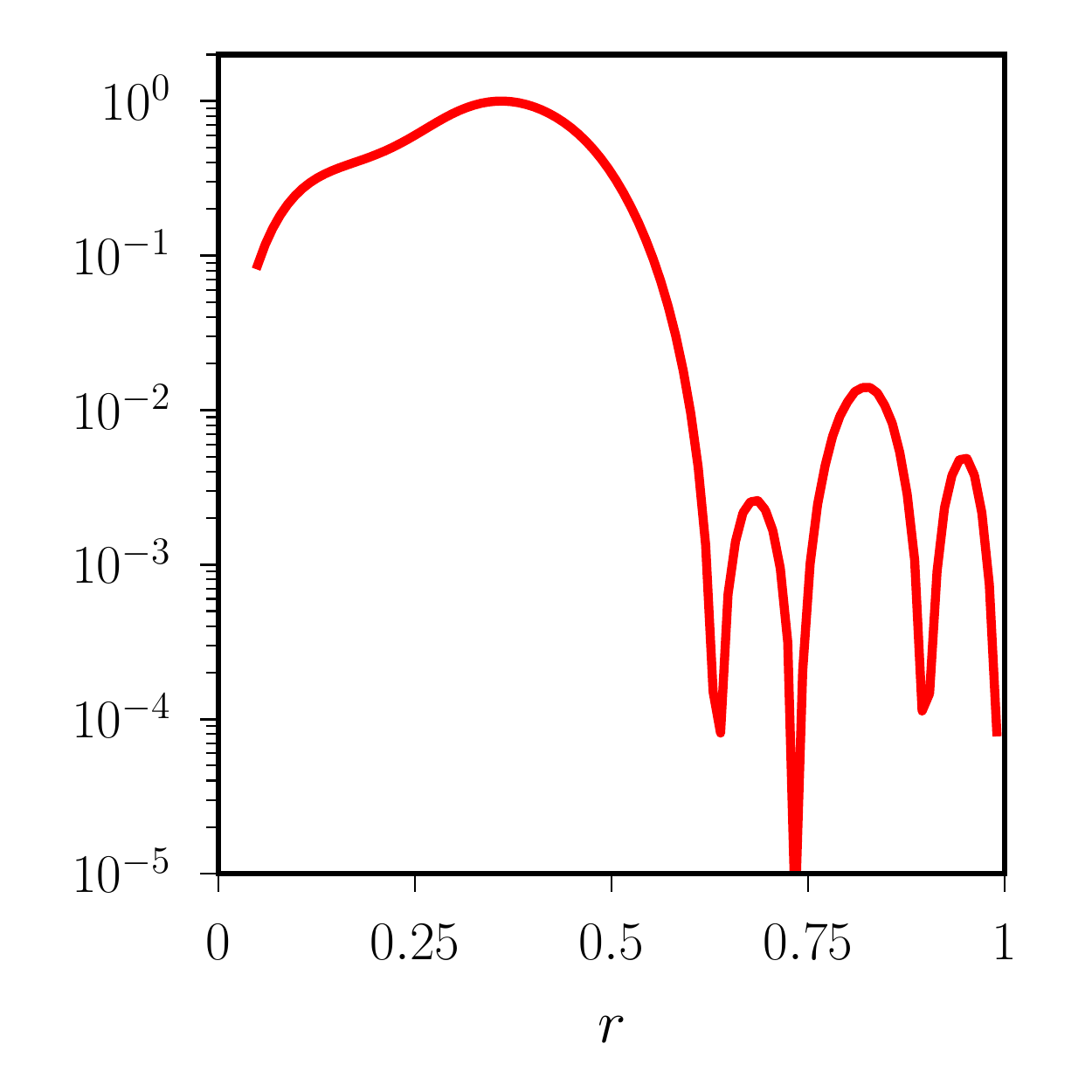} &
    \includegraphics[width=0.38\textwidth]{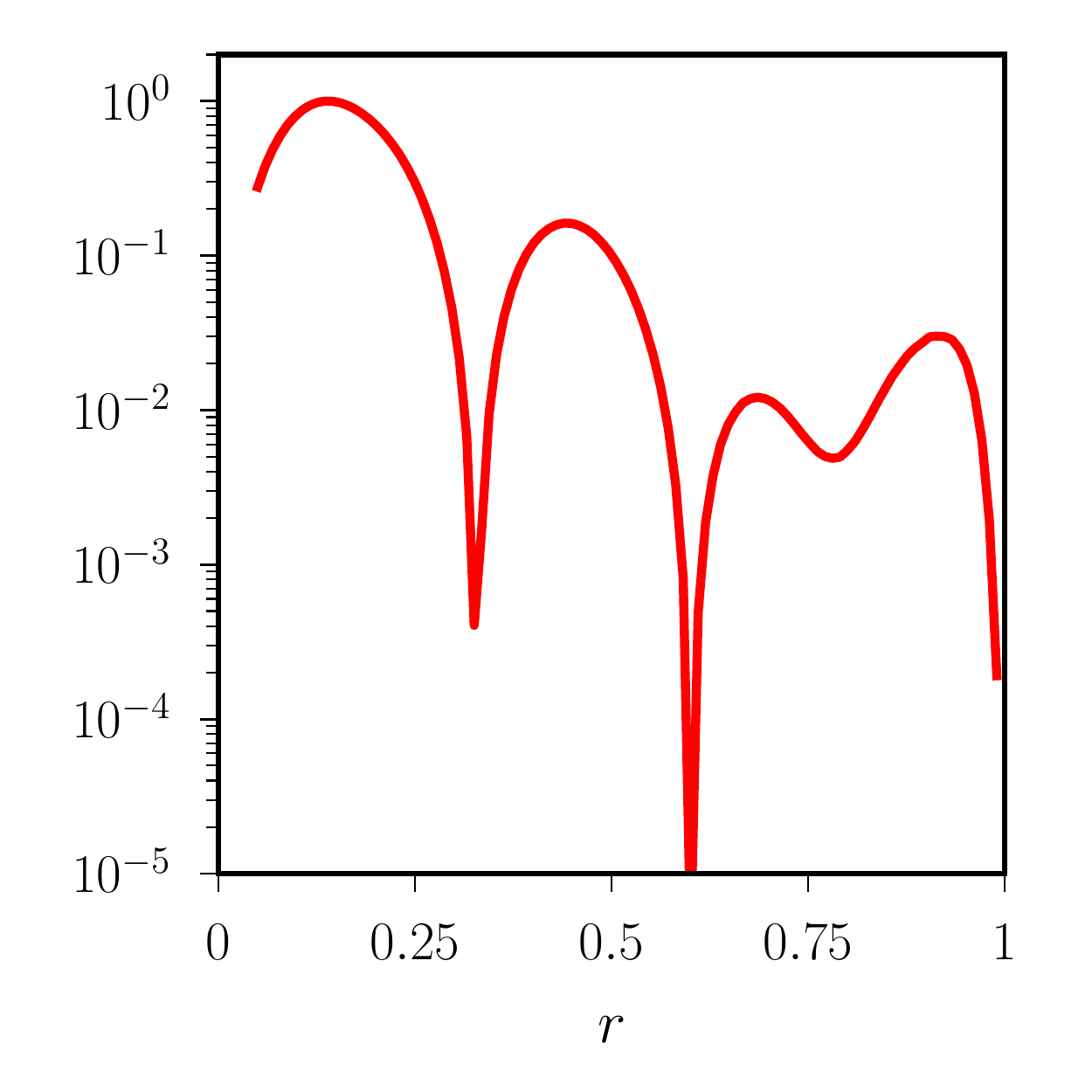} \\
    (a) $|\omega_t|/\wcv = 1.13$ & (b) $|\omega_t|/\wcv = 1.89$ \\
    \includegraphics[width=0.38\textwidth]{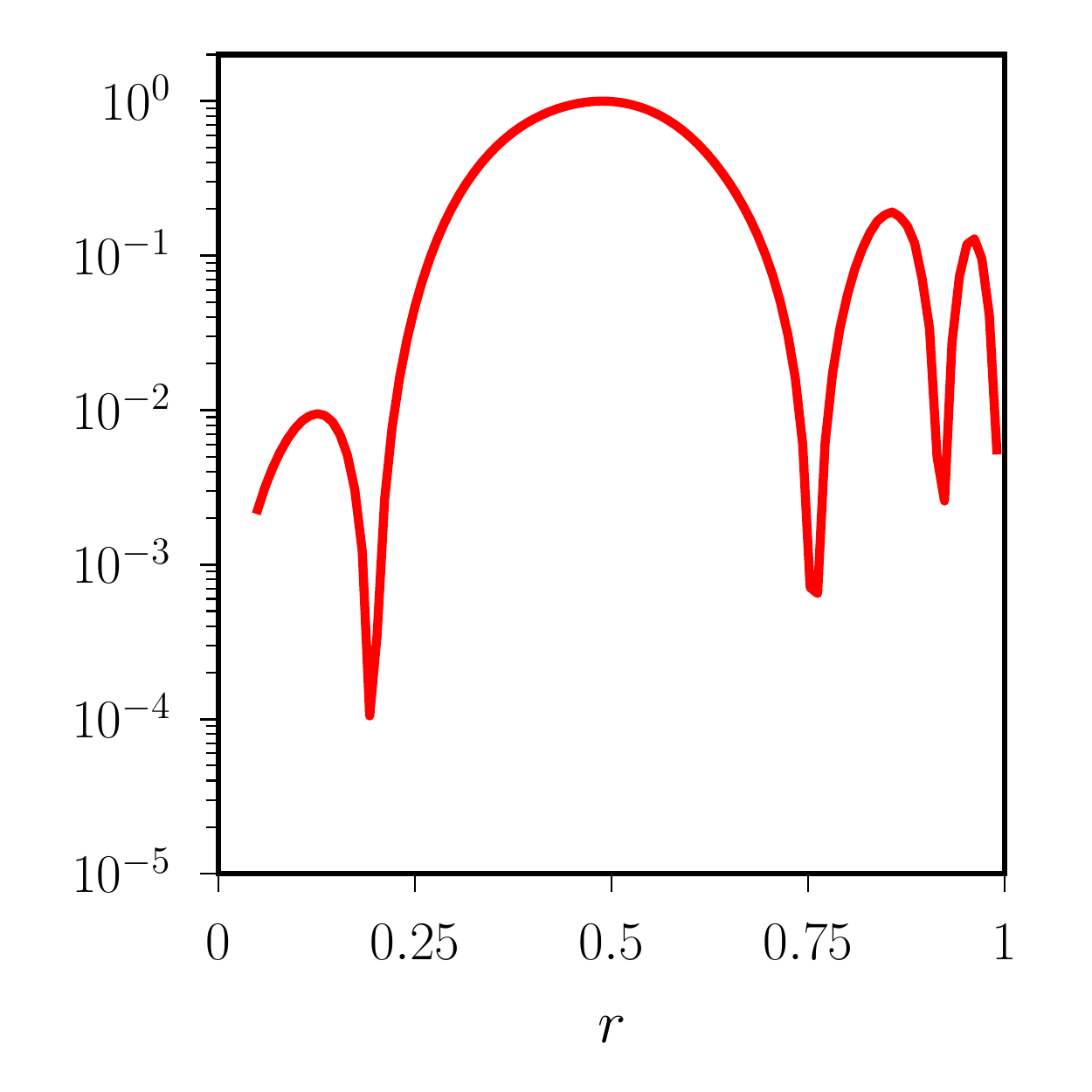} &
    \includegraphics[width=0.38\textwidth]{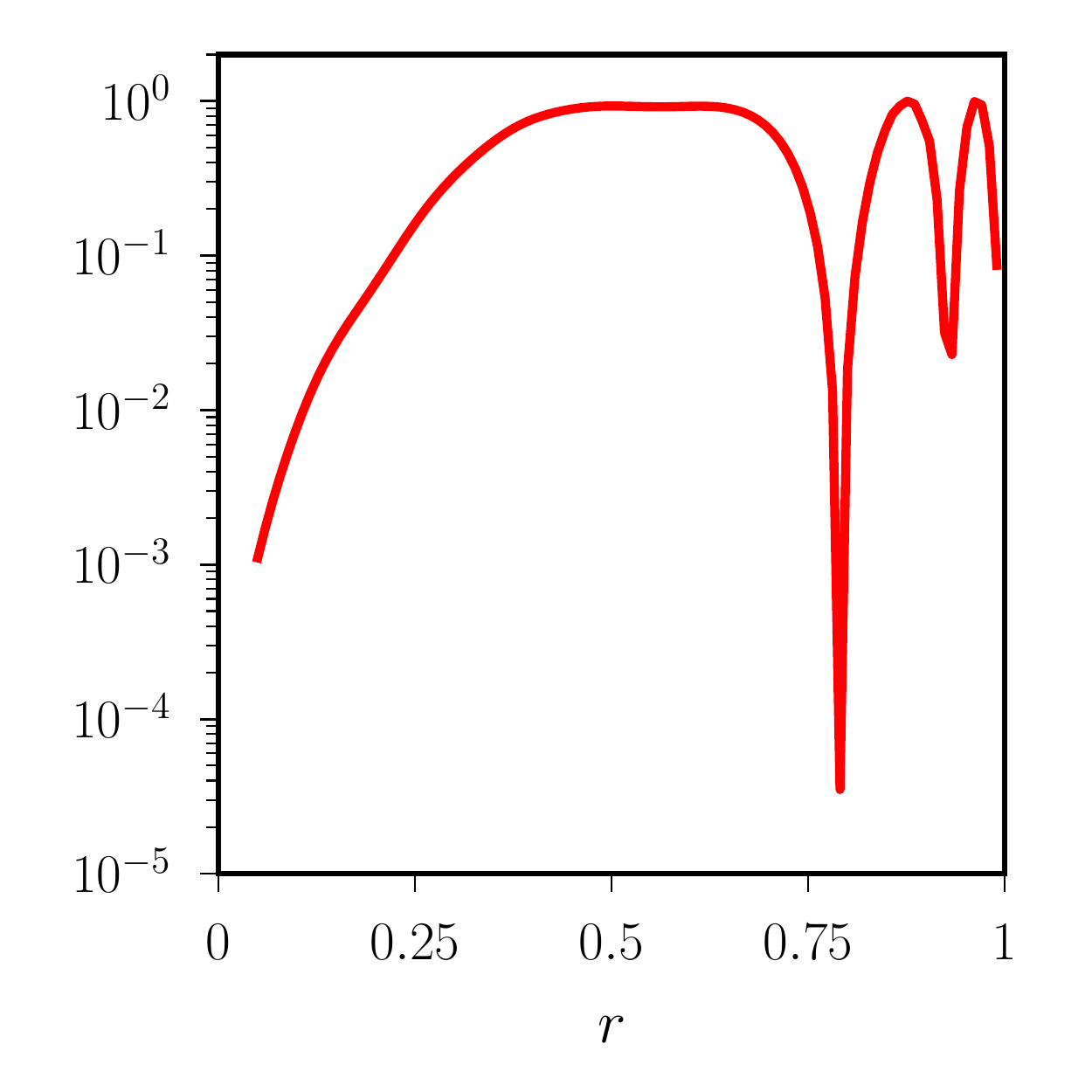} \\
    (c) $|\omega_t|/\wcv = 5.67$ & (d) $|\omega_t|/\wcv = 11.3$ \\
    \end{tabular}
    \caption{Normalized power spectrum of the time-averaged $l=m=0$ component (i.e.~average on a spherical shell) of equation (\ref{eq:eddy}) as a function of radius $r$ in DNS with $Ra=10^6$, $Pr=1$ and $\beta=5 \times 10^{-2}$. These indicate the radii that provide the dominant contribution to $\nu_E$.}
    \label{fig:eddy4}
\end{figure*}

Our DNS have shown that the frequency-reduction of the eddy viscosity is directly correlated with the frequency spectrum of the convection (which is largely unaltered by the tidal flow). 
Outside the dissipation range of the convection, we have recovered the quadratic reduction for frequencies in the Kolmogorov cascade \citep{goldreich1977turbulent}, but for lower frequencies where the frequency spectrum is less steep than the Kolmogorov spectrum, we have found a new frequency reduction that is surprisingly smaller than the linear suppression proposed by \citet{zahn1966marees}.
One could look at the scales that dominate the effective viscosity  to get further physical insight into this problem.  
\citet{zahn1966marees} indeed assumed that the dissipation is dominated by the largest eddies, whereas \cite{goldreich1977turbulent} assumed that the `resonant eddies' dominate the dissipation.

To this end, we illustrate in Fig. \ref{fig:eddy4} the radial dependence of the turbulent viscosity for the illustrative DNS with $Ra=10^6$ and $\beta=5 \times 10^{-2}$ for different tidal frequencies. 
We show the power spectrum (normalized by its maximum value) of the $l=~m=~0$ component (i.e.~the surface-average per shell) of quantity (\ref{eq:eddy}) as a function of the radius $r$. 
Within the anomalous range ($|\omega_t|/\wcv \gtrsim 1$ in Figs \ref{fig:eddy4}a and b), the eddy viscosity is dominated by turbulent eddies deep in the interior.
We also find a significant contribution of the interior eddies in Fig. \ref{fig:eddy4}d, for DNS with much higher frequencies $|\omega_t|/\wcv \geq 10$ (i.e. characterized by the quadratic suppression with negative values), but smaller-scale turbulent interactions are also triggered nearer the surface (except in the outer thin thermal boundary layer).
These radial profiles do not allow us to disentangle easily the length scales that are responsible for the various scaling laws for $\nu_E$. However, they do show a tendency for larger radii to contribute more at high frequencies. 
This trend might be expected if the `resonant eddies' at each radius (with frequencies comparable with $|\omega_t|$) are important, since the convective heat flux increases with radius such that the local convective eddies have larger frequencies nearer the surface. 
However, our simulations do not provide convincing support for this hypothesis \citep[see also in][]{duguid2020tidal}.

\begin{figure}
    \centering
    \includegraphics[width=0.45\textwidth]{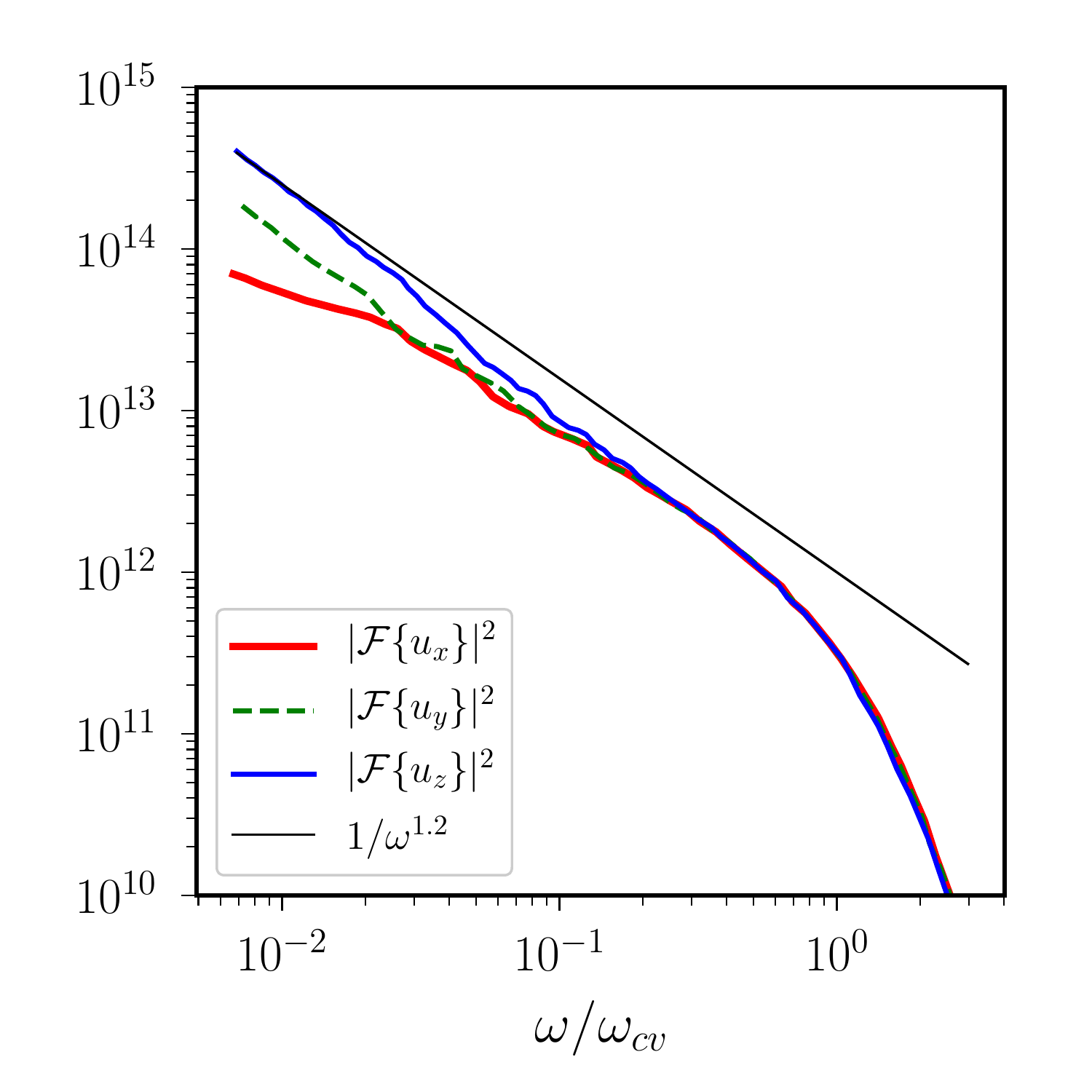}
    \caption{Frequency spectra of the three velocity components $[u_x, u_y, u_z]$ in anelastic DNS, obtained from fig. 3 in \citet{penev2009direct}.}
    \label{fig:penev1}
\end{figure}

In light of our findings, we have revisited the numerical results of \citet{penev2009direct} from an independent viewpoint.
Indeed, they argued that the observed linear scaling for the effective viscosity in their DNS was due to the shallower than Kolmogorov frequency spectrum of the convection.
Hence, one might wonder whether their DNS were subject to $1/\omega^\alpha$ dynamics (as found in our DNS).
We reproduce in Fig. \ref{fig:penev1} the frequency power spectrum of the convective flows in their DNS. 
The spectra are less steep than the expected Kolmogorov spectrum $1/\omega^2$, in broad agreement with the power law $1/\omega^{\delta}$ with\footnote{The exponent $\delta$ given in fig. 3 has a typo in \citet{penev2009direct}, which has been corrected in \citet{penev2011three}.} $\delta\approx 1.2$. 
The latter value is incompatible with our results, since we have always found $1/\omega^\alpha$ power laws with $\alpha < 1$ within the anomalous range. 
Instead, the reduction factor for $\nu_E$ obtained by \citet{penev2009direct} could result from eddies in a turbulent cascade \citep[as in][]{goldreich1977turbulent}, but only if the theoretical scaling for $\nu_E$ is modified to account for spatial spectra with non-standard power exponents ($\neq -5/3$) in the turbulent cascade.

\begin{figure}
    \centering
    \includegraphics[width=0.47\textwidth]{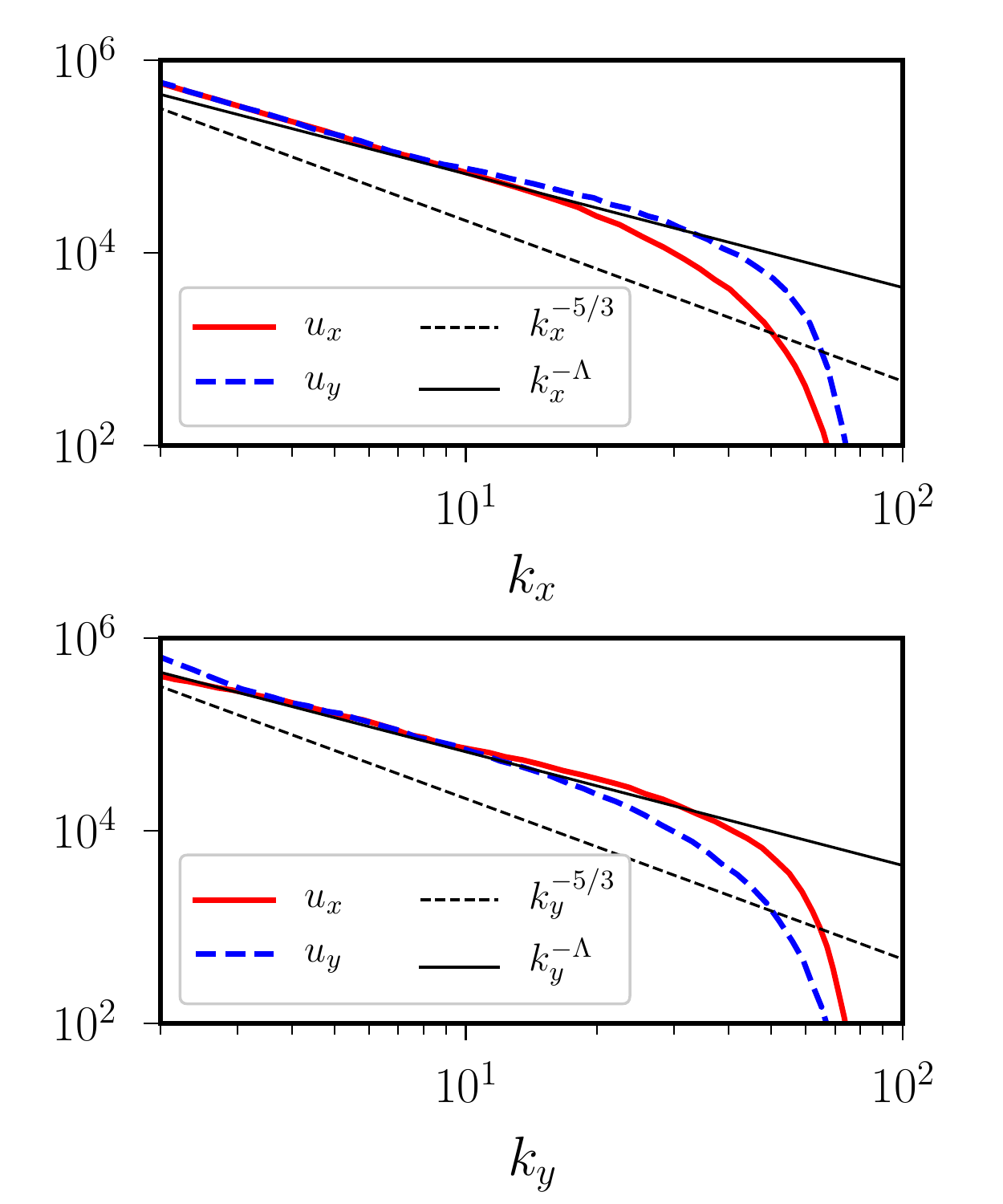}
    \caption{Time-averaged horizontal power spectrum of the velocity components $[u_x, u_y]$ in anelastic DNS \citep[obtained from the top panels of fig. 3 in][]{penev2009direct}, as a function of the horizontal wave numbers $[k_x,k_y]$ in plane-layer geometry.}
    \label{fig:penev2}
\end{figure}

Indeed, a simple predictive theory can be developed for incompressible flows, which relates the power exponent $\delta$ of the frequency spectrum to the power exponent $\Lambda$ of the spatial spectrum of the turbulent kinetic energy, such that \citep{goldman1991orbital} 
\begin{subequations}
\label{eq:KolmorogovMLT}
\begin{equation}
    \Lambda = \frac{3\delta-1}{1+\delta}, \quad \quad \delta = \frac{1+\Lambda}{3-\Lambda}.
    \tag{\theequation a,b}
\end{equation}
\end{subequations}
Standard Kolmogorov turbulence with $\Lambda=5/3$ gives $\delta=2$ \citep{Landau1987Fluid}, as considered by \citet{goldreich1977turbulent}. 
We can then deduce from (\ref{eq:KolmorogovMLT}) that the frequency-reduction of the eddy viscosity is $\nu_E \propto 1/|\omega_t|^\delta$ \citep[see the derivation Appendix A in][]{goldman1991orbital}.
Note that the $1/\omega^\alpha$ spectra observed in our DNS with $\alpha < 1$ cannot be explained by the latter theory, since the spatial exponent $\Lambda$ predicted by (\ref{eq:KolmorogovMLT}a) that is required to match $\delta < 1$ does not agree with the observed spatial spectra in Fig. \ref{fig:speclunperturbed}. 
A non-Kolmogorov ``cascade" with $\Lambda\ne 5/3$ could be produced by scale-dependent buoyant driving or non-negligible viscous damping, and it might also result from anisotropic or inhomogeneous turbulence. 

We show in Fig. \ref{fig:penev2} the time-averaged spatial power spectra of the velocity components reported in \citet{penev2009direct}. 
To be more consistent with the incompressible theory, we have only shown the power spectra of the horizontal velocity components as a function of the horizontal wave numbers $k_x$ and $k_y$ (since their anelastic results could differ more importantly from this simple incompressible theory in the vertical direction, as a result of their adopted density stratification). 
The spatial spectra, which are clearly flatter than the Kolmogorov spectrum (i.e.~with $\Lambda \leq 5/3$), are in good agreement with the power law $k_i^{-\Lambda}$ with the exponent $\Lambda=1.18$ given by expression (\ref{eq:KolmorogovMLT}a) assuming $\delta = 1.2$ (see Fig. \ref{fig:penev1}).

\begin{figure}
    \centering
    \includegraphics[width=0.49\textwidth]{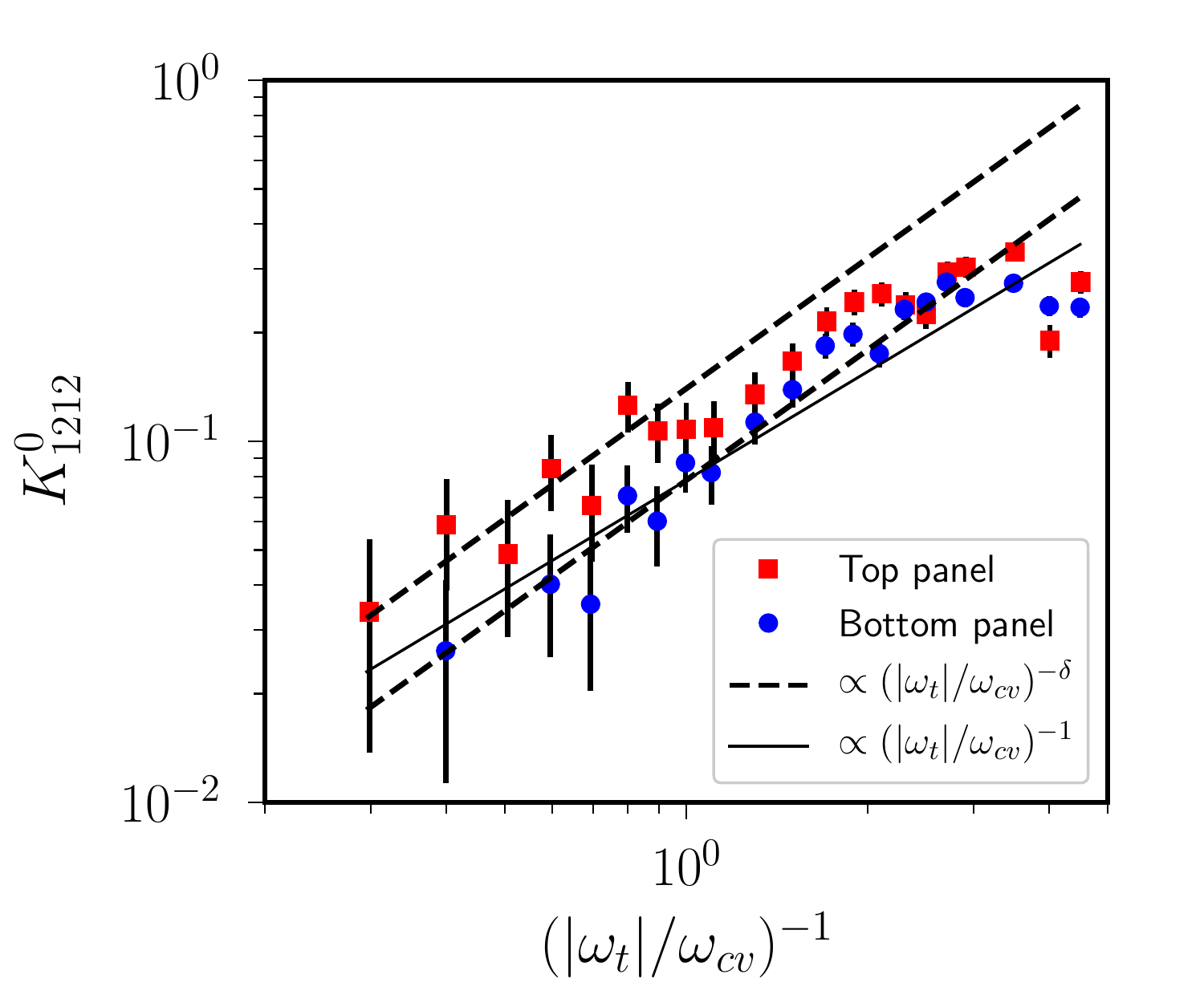}
    \caption{Comparison between theory (\ref{eq:KolmorogovMLT}), linear reduction, and direct measurements of the eddy viscosity (denoted here $K_{1212}^{0}$) in anelastic DNS \citep[extracted from the weak forcing case of fig. 12 in ][]{penev2009direct}. 
    The two data sets (red squares and blue circles) have been computed using two different methods \citep[see details in][]{penev2009direct}.
    Two power laws $K_{1212}^{0} \propto (|\omega_t|/\wcv)^{-\delta}$ with $\delta=1.2$ have been drawn (one for each data set in the range $(|\omega_t|/\wcv)^{-1} \leq 1$).} 
    \label{fig:penevK1212}
\end{figure}

Finally, we reproduce in Fig. \ref{fig:penevK1212} the horizontal effective eddy viscosity coefficient, computed from the DNS of \citet{penev2009direct}, as a function of $(|\omega_t|/\wcv)^{-1}$ using Penev's representation. 
Even if the measurements are subject to relatively large uncertainties, the frequency-reduction of the eddy viscosity in the fast tide range (here $(|\omega_t|/\wcv)^{-1} \leq 1$) is in good agreement with our prediction using equation (\ref{eq:KolmorogovMLT}) assuming $\delta=1.2$.
Moreover, our theory is also more consistent with the fact that the eddies with convective time-scales close to the tidal forcing period were responsible for most of the dissipation in the compressible DNS, as reported by \citet{penev2009direct} (contrary to Zahn's assumption).  
Therefore, the fact that frequency-reduction law reported in \citet{penev2009direct} appears broadly consistent with a linear suppression cannot be taken to conclusively support Zahn's prescription.

To summarize, very different frequency spectra can be generated by turbulent convection, leading to different prescriptions for the frequency-suppression law of the eddy viscosity.
They can manifest in the form of anomalous $1/\omega^\alpha$ power laws for low to intermediate frequencies, such that the frequency-reduction law of the eddy viscosity is expected to be directly correlated with the anomalous frequency spectrum (as reported here).
Additionally, the convection can also exhibit a turbulent cascade that is less steep than the Kolmogorov spectrum \citep[e.g.][]{penev2009direct}, such that the quadratic reduction factor of the eddy viscosity  initially proposed by \citet{goldreich1977turbulent} 
ought to be modified accordingly. 

 \subsection{Astrophysical implications}
\begin{figure}
    \centering
    \includegraphics[width=0.45\textwidth]{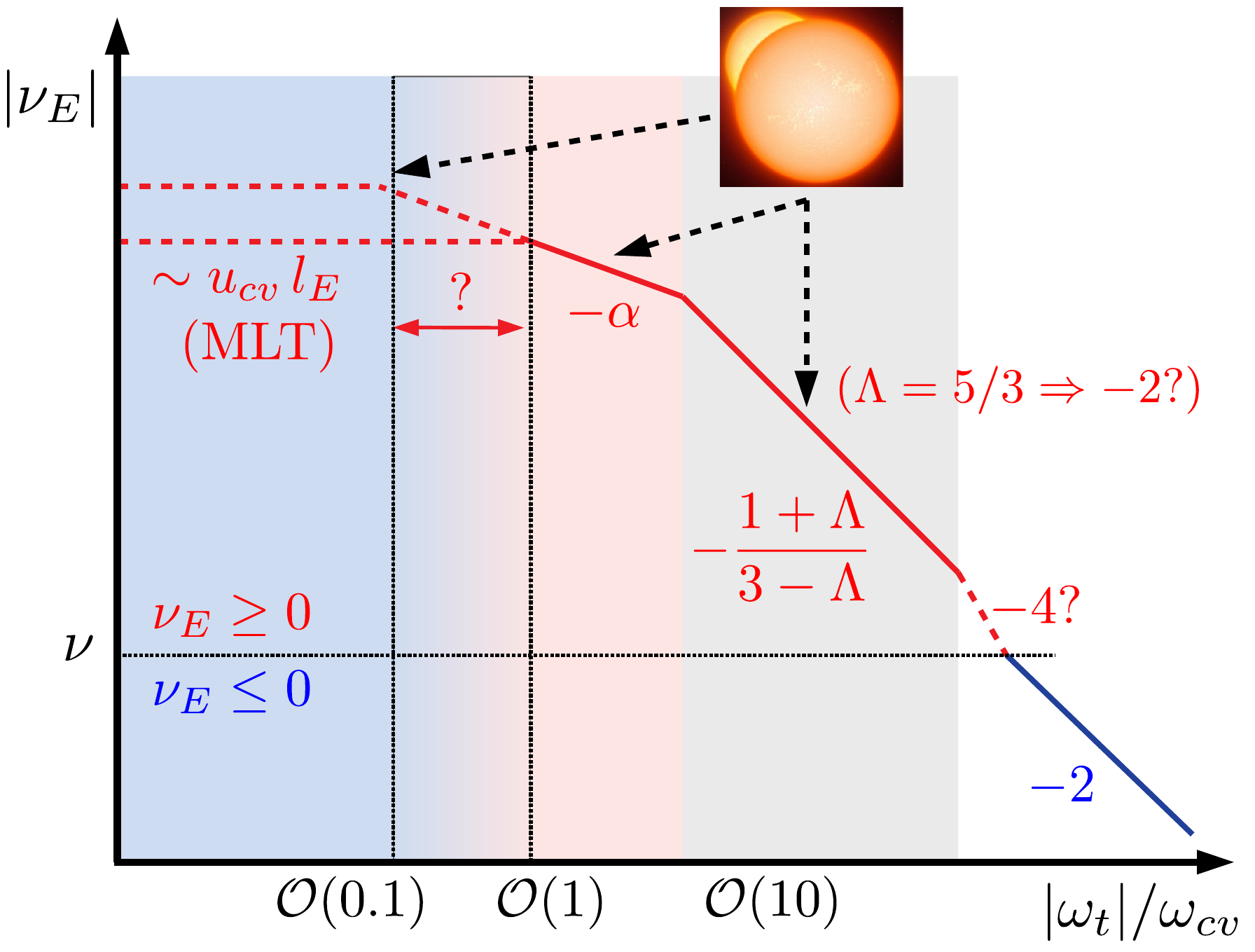}
    \caption{Expected behavior of $\nu_E$ as a function of $|\omega_t|/\wcv$ in turbulent stellar interiors. Laminar viscosity $\nu$, turbulent viscosity $\nu_{E} \sim u_{cv} \, l_E$ (MLT) in the low-frequency regime. $\Lambda$ is the power exponent of the spatial spectrum of the kinetic energy. Background colors refer to Fig. \ref{fig:specFFT}.} 
    \label{fig:cartoon} 
\end{figure}

Our findings indicate that the frequency dependence of the eddy viscosity is surprisingly much more complex than initially proposed by \citet{zahn1966marees} and \citet{goldreich1977turbulent}. We can qualitatively extrapolate our findings to weakly rotating stellar interiors as illustrated in Fig. \ref{fig:cartoon}. 
For very low frequency forcing, standard expectations from MLT \citep[e.g.][]{Spiegel1971} predict the eddy viscosity to scale as $\nu_E \sim u_{cv} \, l_E \propto (Ra/Pr)^{1/2}$ in dimensionless units, independently of the tidal frequency when $|\omega_t|/\wcv \ll 1$. 
The latter scaling is consistent with constant tidal lag-time models \citep[e.g.][]{Alexander1973,hut1981tidal,Eggleton1998}, which are commonly applied in astrophysics. 
However, since this model is only valid for very low tidal frequencies ($|\omega_t|\ll \wcv$),  the constant time-lag model should not be used for the majority of tidal applications, particularly those in which $|\omega_t|\gtrsim \wcv$.

In the presence of fast tides $|\omega_t|/\wcv \geq 1$, the effective viscosity ought to be reduced.
A $1/|\omega_t|^\alpha$ power-law reduction is first expected, with shallow exponents $\alpha < 1$. 
Secondly, for frequencies in a turbulent cascade that is characterized by a power-law spatial spectrum with an arbitrary exponent $\Lambda$, the effective viscosity should be reduced as $\nu_E \propto 1/|\omega_t|^{(1+\Lambda)/(3-\Lambda)}$ \citep{goldman1991orbital}. This gives a quadratic reduction for standard Kolmogorov turbulence \citep[as proposed by][]{goldreich1977turbulent}. 
This quadratic reduction is probably the relevant one in stars and planets \citep[e.g.][]{goldreich1977solar}, but further work is required to assess this hypothesis with more realistic compressible (or anelastic) models. 
Finally, for much higher frequencies, the eddy viscosity may exhibit a quartic reduction $\nu_E \propto 1/|\omega_t|^{4}$ in a narrow transition range towards the dissipation scales of the turbulence, and then a quadratic suppression $|\nu_E| \propto 1/|\omega_t|^{2}$ with possibly negative values for tidal frequencies further into the dissipation range when $|\nu_E| \lesssim \nu$ \citep[see also in][]{ogilvie2012interaction,duguid2020tidal}.

Based on our results, robust quantitative extrapolation is currently challenging beyond the aforementioned qualitative picture. 
The latter two frequency regimes may be not relevant in astrophysics, because they would require very large values of $|\omega_t|/\wcv$.
MLT indeed predicts $\wcv \propto (Ra/Pr)^{1/2}$ in the fully turbulent regime (in broad agreement with our DNS, as shown in the inset panel in Fig. \ref{fig:wcv}). 
For solar-like stars, typical values for the Rayleigh and Prandtl numbers are indeed $Ra = 10^{19}-10^{24}$ and $Pr = 10^{-6}-10^{-4}$ \citep{hanasoge2014quest}, such that the turbulent cascade should extend to much higher frequencies, and the lower bound of the dissipation range should be shifted to $|\omega_t|/\omega_{cv} \gg \mathcal{O}(10)$, compared with our simulations.
Values $\nu_E \gg \nu$ are thus expected in most stellar interiors. 
Negative values $\nu_E \leq 0$ may be theoretically possible in stellar interiors, but very large values of $|\omega_t|$ would probably be required, which are likely to be unrealistic for large-scale tidal flows. 
The turbulent convective damping of the acoustic modes \citep{goldreich1977solar} also provides an indirect viewpoint that may suggest that the observed negative values are not physically relevant.  
Indeed, if the observed correlation between the frequency spectrum of the convection and the frequency-reduction law of the eddy viscosity is generic, then the turbulent cascade should extend until very large frequencies\footnote{The acoustic modes have much larger frequencies than those of tidal forcing.} because a quadratic reduction of positive eddy viscosities is probably required to explain the damping of these modes
\citep[e.g.][]{goldreich1994excitation,Samadi2001}.

The power spectrum observed within the anomalous range may naively appear as a transition between the zero-frequency and the Kolmogorov-like scalings \citep[as in][]{goldman2008effective}. 
However, this is more probably an occurrence of $1/\omega^\alpha$ turbulent noise \citep{niemann2013fluctuations}, which is a robust feature of various turbulent flows \citep[e.g.][]{herault2015experimental,pereira2019noise}. 
This power spectrum may thus exist in turbulent stellar (or planetary) interiors, resulting from the long-term properties of the turbulent flows \citep[according to prior statistical theories, e.g.][]{herault2015stat}. 
We have unfortunately found here power exponents $\alpha < 1$ that vary with $Ra$ and $Pr$ in full spheres (see the slopes in Figs \ref{fig:eddy1}a and \ref{fig:eddy2}, both obtained at $Ra=10^6$), contrary to preliminary findings in plane-layer geometries (with $\alpha \simeq 0.5$, which will be presented elsewhere). 
This indicates an important model-dependence to the anomalous range, and so we cannot currently extrapolate the numerical values of $\alpha$ for very turbulent interiors. 

Our DNS also suggest that $\nu_E$ could be reduced (over its low frequency asymptotic value) for smaller frequencies $|\omega_t|/\omega_{cv} \lesssim 1$ as indicated in Fig. \ref{fig:cartoon}, because the anomalous range may extend until $|\omega_t|/\wcv \simeq \mathcal{O}(0.1)$ or perhaps below (as observed in the various frequency spectra).
Yet, since computations of the low-frequency spectrum of turbulent flows are very challenging, we have been unable to directly measure the eddy viscosity within the low-frequency regime, and we do not have very reliable estimates of the transition values between the two regimes for very turbulent stellar interiors.

\begin{figure}
    \centering
    \includegraphics[width=0.43\textwidth]{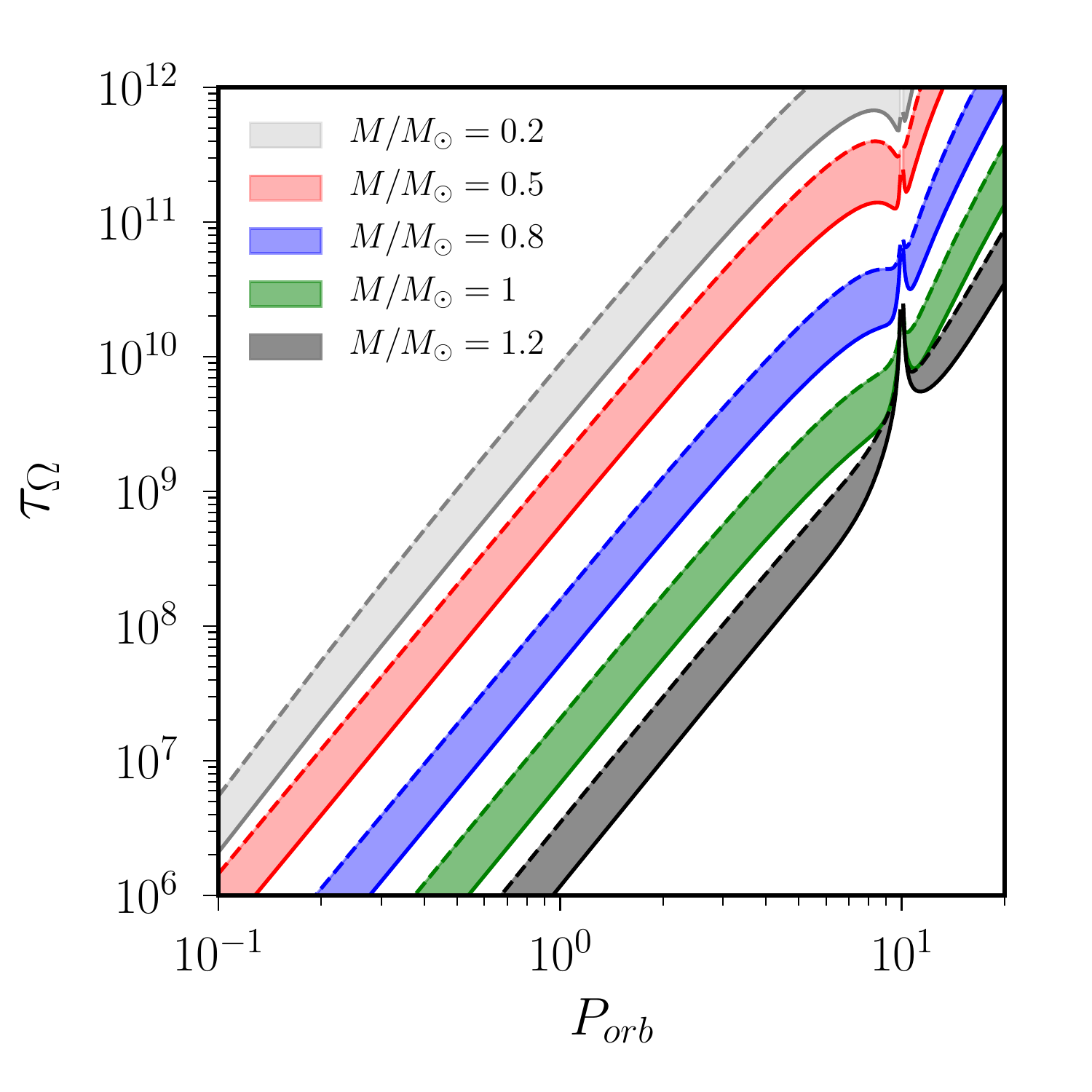}
    \caption{Synchronization time-scale $\tau_\Omega$ (in years), as a function of $P_{orb}$ (in days), due to convective damping of the equilibrium tide \citep[e.g.][]{terquem1998tidal,ogilvie2014tidal}, for a star of mass $M$ with the initial spin period $P_s = 10$ d. The companion has a fixed mass $M_2 = M_\odot$. Viscosity prescription based on Fig. \ref{fig:cartoon} with $0.5 \leq \alpha \leq 1$ within the anomalous range $0.3 < |\omega_t|/\omega_{cv} < 3$, and a quadratic reduction when $|\omega_t|/\omega_{cv} \geq 3$ (i.e. with $\Lambda=5/3$). Solid lines (respectively dashed lines) have been computed with $\alpha=0.5$ (respectively $\alpha=1$).}
    \label{fig:synchro}
\end{figure}

To illustrate one of the uncertainties in applying our results, we briefly explore how the theoretical time-scales for binary spin synchronization are affected by changes in the slope $\alpha$ of the anomalous regime. 
To do so, we consider a continuous piece-wise power-law profile for $\nu_E$ based on our simulations (as illustrated in Fig. \ref{fig:cartoon}).
We adopt $\nu_E = u_{cv} \, l_E$ for $\omega_t/\omega_{cv}\leq0.3$, then $\nu_E\propto 1/|\omega_t|^{\alpha}$ for $0.3 < |\omega_t|/\omega_{cv} < 3$, and finally $\nu_E \propto 1/|\omega_t|^{2}$ when $|\omega_t|/\omega_{cv}\geq 3$ (discarding the possible negative values of $\nu_E$).
We consider the values $0.5 \leq \alpha \leq 1$ that span our simulations. 
We use main-sequence stellar models computed with MESA (see Appendix \ref{MESA} for further details), where $u_{cv}$ and $l_E$ are here the convective velocity and mixing length that vary with stellar radius, and $\omega_{cv}=u_{cv}/l_E$. 
We calculate the correct equilibrium tide in convective regions  \citep{terquem1998tidal,ogilvie2014tidal}, which differs from the commonly-adopted but strictly incorrect equilibrium tide of \citet{zahn1989tidal}, and then compute the dissipation integral. 
We thus obtain a tidal quality factor $Q'_\mathrm{eq}$, from which the time-scale for tidal synchronization of the stellar spin of the primary star interacting with a companion of mass $M_2$ is \citep[after correcting a typographical error in formula (7) of][]{vidal2020turbulent}
\begin{equation}
    \label{tauOm}
    \tau_\Omega = \frac{2 Q'_\mathrm{eq}}{9\pi  r_g^2}\left(\frac{M+M_2}{M_2}\right)^2\frac{P_{orb}^4}{P_{dyn}^2P_s},
\end{equation}
where $r_g^2$ is the dimensionless squared radius of gyration, $P_{dyn}=2\pi/(GM/R^3)^{1/2}$ is the dynamical time-scale, $P_{orb}=2\pi/\Omega_{orb}$ is the orbital period, and $P_s=2\pi/\Omega_s$ is the (initial) spin period. 
We show in Fig. \ref{fig:synchro} the results for $\tau_\Omega$ as a function of $P_{orb}$ with $M_2=M_\odot$ (where $M_\odot$ is the solar mass) and $P_s=10$ d in each case, for a range of main-sequence stellar models with masses $M/M_\odot\in [0.2,0.5,0.8,1.0,1.2]$ that correspond to the stellar ages $[2.9,3.3,2.6,4.7,2.9]$ Gyr. 
This shows that for an anomalous regime spanning a decade in frequency, uncertainties in $\alpha$ only affect $\tau_\Omega$ by a factor of two or three (except near spin-orbit synchronization at $P_{orb}=P_s= 10$). 
On the other hand, if the anomalous range is much wider, uncertainties in $\alpha$ could have more important effects on $\tau_\Omega$ (not shown).

Finally, the power spectrum of the turbulent cascade is also uncertain. 
Kolmogorov spectra have been robustly reported for Boussinesq convection \citep{kumar2018applicability}, but compressible convection \citep[e.g.][]{penev2011three,horst2020fully} may display different non-Kolmogorov spectra (depending on the convection setup). 
Further work is required to characterize the frequency spectrum of more realistic stellar convection so that we can robustly apply our results to astrophysical tidal evolution.

\section{Concluding remarks}
\label{sec:ccl}
In this paper, we have revisited numerically the long-standing controversy regarding the interaction between equilibrium tidal flows and turbulent convection.
We have conducted DNS of thermal convection within an idealized global model of a fully-convective fluid body, which is a simple analogue of a low-mass star or core-less giant planet, to measure the turbulent viscosity $\nu_E$ acting on the large-scale equilibrium tidal flow. 

Our results have highlighted that quantifying the efficiency of tidal dissipation resulting from this mechanism is more complex than was previously believed. Indeed, we have found that neither the often-disputed linear \citep{zahn1966marees} or quadratic \citep{goldreich1977turbulent} scaling laws for the turbulent viscosity are generally valid for tidal frequencies $|\omega_t|$ that exceed the dominant convective turnover frequency $\wcv$.
Instead, we have demonstrated that the frequency-reduction law of the eddy viscosity is correlated with the frequency spectrum of the unperturbed convection, and we have obtained here various scaling laws in our DNS.

The eddy viscosity is first reduced as $\nu_E \propto 1/|\omega_t|^\alpha$ 
for tidal frequencies below those in the turbulent cascade, with shallow exponents $\alpha < 1$. 
Then, for frequencies in a turbulent Kolmogorov-like cascade with the spatial power exponent $\Lambda$, we have consistently combined our numerical findings with Penev's previous results to show that $\nu_E \propto 1/|\omega_t|^{(1+\Lambda)/(3-\Lambda)}$. For the standard Kolmogorov cascade with $\Lambda=5/3$, which is probably the relevant one over a broad range of scales in stars or planets \citep[as supported by observations of solar acoustic modes, e.g.][]{goldreich1977solar,Samadi2001}, this leads to $\nu_E \propto 1/|\omega_t|^2$ \citep{goldreich1977turbulent}. However, further work is required to explore the robustness of this scaling in more realistic (compressible or anelastic) models.
Our results finally support the universality of the quadratic reduction law $|\nu_E| \propto |\omega_t|^{-2}$ for very high frequencies in the dissipation range of the convection, which is consistent with asymptotic predictions when $|\omega_t|/\omega_{cv} \gg 1$ \citep{ogilvie2012interaction,duguid2020tidal}.

Our findings have important consequences for interpreting astrophysical observations such as those that constrain tidal synchronization and circularization of main-sequence binaries 
\citep[e.g.][]{meibom2005robust,meibom2006observational,van2016orbital,lurie2017tidal,triaud2017eblm} and the circularization of evolved stars \citep{verbunt1995tidal,Beck2018,price2018binary}. 
Indeed, it appears that a fundamental knowledge of stellar convection is required before we can be confident in modeling the tidal evolution of astrophysical systems due to this mechanism. 
Hence, further work is required to understand the properties of more realistic convection models in the presence of oscillatory tidal flows.
The transitions between the various regimes observed in our DNS remain for instance poorly constrained, since we have necessarily adopted simulation parameters that are far removed from their astrophysical values, and so should be further explored in more realistic models of stellar convection.
The anomalous $1/\omega^\alpha$ spectrum should be also further investigated as a function of $Ra/Pr$, as well as the slope of power spectrum of the turbulent cascade, which would be very challenging numerically in more turbulent setups. 
Astrophysical extrapolations also employ crude applications of MLT to the low-frequency regime, which is known not to be accurate in detail \citep[e.g.][]{goldman2008effective} and departures from MLT have been found in DNS of compressible convection \citep[e.g.][]{anders2019entropy}. 
Hence, MLT predictions should be carefully compared to more turbulent DNS of convection. 

We have considered only circular orbits in this paper, but different tidal components generally coexist \citep[e.g. for eccentric orbits, see in][]{IvPap2004,vick2020tidal} and they could be damped at different rates \citep[e.g.][]{lai2012tidal}. We have also neglected dynamical tides \citep[e.g.][]{OL2007}, although their interaction with convection may be important when inertial waves are excited.
Indeed, tidally-excited inertial waves (restored by Coriolis forces) may be the key driver of binary circularization and synchronization in sufficiently rapidly rotating stars
\citep[e.g.][]{OL2007,GoodmanLackner2009,IvPap2013,favier2014non}. 

Finally, note that our simple physical picture should remain qualitatively valid in weakly rotating interiors (i.e. slow rotators). However, rapid rotation is known to strongly affect convection-driven turbulence, as reported in DNS of plane-layer \citep[e.g.][]{barker2014theory,Currie2020} and spherical convection \citep[e.g.][]{kaplan2017subcritical,guervilly2019turbulent}, and it has also been proposed that it could modify the effective viscosity  \citep{mathis2016impact}. 
Further work is required to explore rapidly rotating convection, which might be relevant for giant planets or young rapidly rotating stars. Since the frequency spectrum of the convection could be strongly impacted by rapid global rotation, the interactions between tidal flows and convection is worth investigating for these applications. Non-linear tidal flows can also be triggered in rapidly rotating interiors for sufficiently large tidal deformations \citep[such as the elliptical (tidal) instability, e.g.][]{barker2016non,vidal2017inviscid}, which could enhance tidal dissipation for the shortest orbital periods
\citep{barker2016nonb,vidal2018magnetic,vidal2019binaries}. 
Understanding the interplay of these flows with convection also deserves future work. 

\section*{Acknowledgements}
We acknowledge the  referee, Adam S. Jermyn, for a prompt report that helped us to significantly improve the clarity of the paper. 
This work was funded by STFC Grant ST/R00059X/1.
This work used the DiRAC@Durham facility managed by the Institute for Computational Cosmology on behalf of the STFC DiRAC HPC Facility. 
The equipment was funded by BEIS capital funding via STFC capital grants ST/P002293/1, ST/R002371/1 and ST/S002502/1, Durham University and STFC operations grant ST/R000832/1. 
DiRAC is part of the National e-Infrastructure. 
Some DNS were also performed using the UKMHD1-UKMHD2-UKMHD3 allocations on the DiRAC Data Intensive service at Leicester, operated by the University of Leicester IT Services, which forms part of the STFC DiRAC High-Performance Computing (HPC) Facility.
Post-processing was performed on ARC4, part of the HPC facilities at the University of Leeds (UK). 

\section*{Data availability}
The python scripts and post-processed data underlying the figures are provided as supplementary materials. 
Data underlying the Figs \ref{fig:penev1}-\ref{fig:penevK1212} have been extracted from the original figures of \citet{penev2009direct} using the online tool \url{https://apps.automeris.io/wpd/}. 
The spherical harmonics analysis has been performed using the open-source library \textsc{shtns} \citep{schaeffer2013efficient}, available at \url{https://bitbucket.org/nschaeff/shtns/}.
The DNS underlying this article can be replicated using the open-source code \textsc{Nek5000}, available at \url{https://nek5000.mcs.anl.gov/}.




\bibliographystyle{mnras}
\bibliography{main} 

\begin{thebibliography}{}
\makeatletter
\relax
\def\mn@urlcharsother{\let\do\@makeother \do\$\do\&\do\#\do\^\do\_\do\%\do\~}
\def\mn@doi{\begingroup\mn@urlcharsother \@ifnextchar [ {\mn@doi@}
  {\mn@doi@[]}}
\def\mn@doi@[#1]#2{\def\@tempa{#1}\ifx\@tempa\@empty \href
  {http://dx.doi.org/#2} {doi:#2}\else \href {http://dx.doi.org/#2} {#1}\fi
  \endgroup}
\def\mn@eprint#1#2{\mn@eprint@#1:#2::\@nil}
\def\mn@eprint@arXiv#1{\href {http://arxiv.org/abs/#1} {{\tt arXiv:#1}}}
\def\mn@eprint@dblp#1{\href {http://dblp.uni-trier.de/rec/bibtex/#1.xml}
  {dblp:#1}}
\def\mn@eprint@#1:#2:#3:#4\@nil{\def\@tempa {#1}\def\@tempb {#2}\def\@tempc
  {#3}\ifx \@tempc \@empty \let \@tempc \@tempb \let \@tempb \@tempa \fi \ifx
  \@tempb \@empty \def\@tempb {arXiv}\fi \@ifundefined
  {mn@eprint@\@tempb}{\@tempb:\@tempc}{\expandafter \expandafter \csname
  mn@eprint@\@tempb\endcsname \expandafter{\@tempc}}}

\bibitem[\protect\citeauthoryear{Alexander}{Alexander}{1973}]{Alexander1973}
Alexander M.~E.,  1973, \mn@doi [\apss] {https://doi.org/10.1007/BF00645172},
  23, 459

\bibitem[\protect\citeauthoryear{Anders, Lecoanet  \& Brown}{Anders
  et~al.}{2019}]{anders2019entropy}
Anders E.~H.,  Lecoanet D.,   Brown B.~P.,  2019, \mn@doi [ApJ]
  {https://doi.org/10.3847/1538-4357/ab3644}, 884, 65

\bibitem[\protect\citeauthoryear{Barker}{Barker}{2016}]{barker2016nonb}
Barker A.~J.,  2016, \mn@doi [MNRAS] {https://doi.org/10.1093/mnras/stw702},
  459, 939

\bibitem[\protect\citeauthoryear{Barker \& Lithwick}{Barker \&
  Lithwick}{2013}]{barker2013non}
Barker A.~J.,  Lithwick Y.,  2013, \mn@doi [MNRAS]
  {https://doi.org/10.1093/mnras/stt1561}, 435, 3614

\bibitem[\protect\citeauthoryear{Barker, Dempsey  \& Lithwick}{Barker
  et~al.}{2014}]{barker2014theory}
Barker A.~J.,  Dempsey A.~M.,   Lithwick Y.,  2014, \mn@doi [ApJ]
  {https://doi.org/10.1088/0004-637X/791/1/13}, 791, 13

\bibitem[\protect\citeauthoryear{Barker, Braviner  \& Ogilvie}{Barker
  et~al.}{2016}]{barker2016non}
Barker A.~J.,  Braviner H.~J.,   Ogilvie G.~I.,  2016, \mn@doi [MNRAS]
  {https://doi.org/10.1093/mnras/stw701}, 459, 924

\bibitem[\protect\citeauthoryear{Beck, Mathis, Gallet, Charbonnel, Benbakoura
  \& Garc\'ia}{Beck et~al.}{2018}]{Beck2018}
Beck P.~G.,  Mathis S.,  Gallet F.,  Charbonnel C.,  Benbakoura M.,   Garc\'ia
  R. A .and~do Nascimento J.-D.,  2018, \mn@doi [\mnras]
  {https://doi.org/10.1093/mnrasl/sly114}, 479, L123

\bibitem[\protect\citeauthoryear{Braviner}{Braviner}{2016}]{braviner2016stellar}
Braviner H.~J.,  2016, PhD thesis, Univ. Cambridge

\bibitem[\protect\citeauthoryear{Currie, {Barker}, Lithwick  \&
  Browning}{Currie et~al.}{2020}]{Currie2020}
Currie L.~K.,  {Barker} A.~J.,  Lithwick Y.,   Browning M.~K.,  2020, \mn@doi
  [\mnras] {https://doi.org/10.1093/mnras/staa372}, 493, 5233

\bibitem[\protect\citeauthoryear{Duguid, Barker  \& Jones}{Duguid
  et~al.}{2020}]{duguid2020tidal}
Duguid C.~D.,  Barker A.~J.,   Jones C.~A.,  2020, \mn@doi [\mnras]
  {https://doi.org/10.1093/mnras/stz2899}, 491, 923

\bibitem[\protect\citeauthoryear{Eggleton, Kiseleva  \& Hut}{Eggleton
  et~al.}{1998}]{Eggleton1998}
Eggleton P.~P.,  Kiseleva L.~G.,   Hut P.,  1998, \mn@doi [\apj]
  {https://doi.org/10.1086/305670}, 499, 853

\bibitem[\protect\citeauthoryear{Favier, Barker, Baruteau  \& Ogilvie}{Favier
  et~al.}{2014}]{favier2014non}
Favier B.,  Barker A.~J.,  Baruteau C.,   Ogilvie G.~I.,  2014, \mn@doi [MNRAS]
  {https://doi.org/10.1093/mnras/stu003}, 439, 845

\bibitem[\protect\citeauthoryear{Fischer, Loth, Lee, Lee, Smith  \&
  Bassiouny}{Fischer et~al.}{2007}]{fischer2007simulation}
Fischer P.~F.,  Loth F.,  Lee S.~E.,  Lee S.-W.,  Smith D.~S.,   Bassiouny
  H.~S.,  2007, \mn@doi [Comput. Methods Appl. Mech. Eng.]
  {https://doi.org/10.1016/j.cma.2006.10.015}, 196, 3049

\bibitem[\protect\citeauthoryear{Gastine, Wicht  \& Aubert}{Gastine
  et~al.}{2016}]{gastine2016scaling}
Gastine T.,  Wicht J.,   Aubert J.,  2016, \mn@doi [J. Fluid Mech.]
  {https://doi.org/10.1017/jfm.2016.659}, 808, 690

\bibitem[\protect\citeauthoryear{Goldman}{Goldman}{2008}]{goldman2008effective}
Goldman I.,  2008, \mn@doi [Astron. Nachr.]
  {https://doi.org/10.1002/asna.200811016}, 329, 762

\bibitem[\protect\citeauthoryear{Goldman \& Mazeh}{Goldman \&
  Mazeh}{1991}]{goldman1991orbital}
Goldman I.,  Mazeh T.,  1991, \mn@doi [\apj] {10.1086/170275}, 376, 260

\bibitem[\protect\citeauthoryear{Goldreich \& Keeley}{Goldreich \&
  Keeley}{1977}]{goldreich1977solar}
Goldreich P.,  Keeley D.~A.,  1977, \mn@doi [\apj] {10.1086/155005}, 211, 934

\bibitem[\protect\citeauthoryear{Goldreich \& Nicholson}{Goldreich \&
  Nicholson}{1977}]{goldreich1977turbulent}
Goldreich P.,  Nicholson P.~D.,  1977, \mn@doi [Icarus]
  {https://doi.org/10.1016/0019-1035(77)90163-4}, 30, 301

\bibitem[\protect\citeauthoryear{Goldreich, Murray  \& Kumar}{Goldreich
  et~al.}{1994}]{goldreich1994excitation}
Goldreich P.,  Murray N.,   Kumar P.,  1994, \mn@doi [\apj] {10.1086/173904},
  424, 466

\bibitem[\protect\citeauthoryear{Gonczi}{Gonczi}{1982}]{gonczi1982local}
Gonczi G.,  1982, A\&A, 110, 1

\bibitem[\protect\citeauthoryear{Goodman \& Lackner}{Goodman \&
  Lackner}{2009}]{GoodmanLackner2009}
Goodman J.,  Lackner C.,  2009, \mn@doi [\apj]
  {https://doi.org/10.1088/0004-637X/696/2/2054}, 696, 2054

\bibitem[\protect\citeauthoryear{Goodman \& Oh}{Goodman \&
  Oh}{1997}]{goodman1997fast}
Goodman J.,  Oh S.~P.,  1997, \mn@doi [\apj] {https://doi.org/10.1086/304505},
  486, 403

\bibitem[\protect\citeauthoryear{Guervilly, Cardin  \& Schaeffer}{Guervilly
  et~al.}{2019}]{guervilly2019turbulent}
Guervilly C.,  Cardin P.,   Schaeffer N.,  2019, \mn@doi [Nature]
  {https://doi.org/10.1038/s41586-019-1301-5}, 570, 368

\bibitem[\protect\citeauthoryear{Hanasoge \& Sreenivasan}{Hanasoge \&
  Sreenivasan}{2014}]{hanasoge2014quest}
Hanasoge S.~M.,  Sreenivasan K.,  2014, \mn@doi [Sol. Phys.]
  {https://doi.org/10.1007/s11207-014-0471-4}, 289, 3403

\bibitem[\protect\citeauthoryear{Herault, P{\'e}tr{\'e}lis  \& Fauve}{Herault
  et~al.}{2015a}]{herault2015experimental}
Herault J.,  P{\'e}tr{\'e}lis F.,   Fauve S.,  2015a, \mn@doi [Europhys. Lett.]
  {https://doi.org/10.1209/0295-5075/111/44002}, 111, 44002

\bibitem[\protect\citeauthoryear{Herault, P{\'e}tr{\'e}lis  \& Fauve}{Herault
  et~al.}{2015b}]{herault2015stat}
Herault J.,  P{\'e}tr{\'e}lis F.,   Fauve S.,  2015b, \mn@doi [J. Stat. Phys.]
  {https://doi.org/10.1007/s10955-015-1408-5}, 161, 1379

\bibitem[\protect\citeauthoryear{Horst, Edelmann, Andrassy, Roepke, Bowman,
  Aerts  \& Ratnasingam}{Horst et~al.}{2020}]{horst2020fully}
Horst L.,  Edelmann P. V.~F.,  Andrassy R.,  Roepke F.~K.,  Bowman D.~M.,
  Aerts C.,   Ratnasingam R.~P.,  2020, \mn@doi [arXiv preprint]
  {https://arxiv.org/abs/2006.03011}

\bibitem[\protect\citeauthoryear{Hut}{Hut}{1981}]{hut1981tidal}
Hut P.,  1981, A\&A, 99, 126

\bibitem[\protect\citeauthoryear{Ivanov \& Papaloizou}{Ivanov \&
  Papaloizou}{2004}]{IvPap2004}
Ivanov P.~B.,  Papaloizou J. C.~B.,  2004, \mn@doi [\mnras]
  {https://doi.org/10.1111/j.1365-2966.2004.08136.x}, 353, 1161

\bibitem[\protect\citeauthoryear{{Ivanov}, {Papaloizou}  \& {Chernov}}{{Ivanov}
  et~al.}{2013}]{IvPap2013}
{Ivanov} P.~B.,  {Papaloizou} J.~C.~B.,   {Chernov} S.~V.,  2013, \mn@doi
  [\mnras] {https://doi.org/10.1093/mnras/stt595}, 432, 2339

\bibitem[\protect\citeauthoryear{Kaplan, Schaeffer, Vidal  \& Cardin}{Kaplan
  et~al.}{2017}]{kaplan2017subcritical}
Kaplan E.~J.,  Schaeffer N.,  Vidal J.,   Cardin P.,  2017, \mn@doi [Phys. Rev.
  Lett.] {https://dx.doi.org/10.1103/PhysRevLett.119.094501}, 119, 094501

\bibitem[\protect\citeauthoryear{Kirk et~al.}{Kirk
  et~al.}{2016}]{kirk2016kepler}
Kirk B.,  et~al., 2016, \mn@doi [\apj]
  {https://doi.org/10.3847/0004-6256/151/3/68}, 151, 68

\bibitem[\protect\citeauthoryear{Kumar \& Verma}{Kumar \&
  Verma}{2018}]{kumar2018applicability}
Kumar A.,  Verma M.~K.,  2018, \mn@doi [RSOS]
  {https://doi.org/10.1098/rsos.172152}, 5, 172152

\bibitem[\protect\citeauthoryear{Lai}{Lai}{2012}]{lai2012tidal}
Lai D.,  2012, \mn@doi [\mnras]
  {https://doi.org/10.1111/j.1365-2966.2012.20893.x}, 423, 486

\bibitem[\protect\citeauthoryear{Lai, Rasio  \& Shapiro}{Lai
  et~al.}{1993}]{lai1993ellipsoidal}
Lai D.,  Rasio F.~A.,   Shapiro S.~L.,  1993, \mn@doi [\apjs] {10.1086/191822},
  88, 205

\bibitem[\protect\citeauthoryear{Landau \& Lifshitz}{Landau \&
  Lifshitz}{1987}]{Landau1987Fluid}
Landau L.~D.,  Lifshitz E.~M.,  1987, Theoretical {Physics}. {Fluid}
  {Mechanics}, second edn.
Pergamon Press, Oxford

\bibitem[\protect\citeauthoryear{Le~Bars, C{\'e}bron  \& Le~Gal}{Le~Bars
  et~al.}{2015}]{le2015flows}
Le~Bars M.,  C{\'e}bron D.,   Le~Gal P.,  2015, \mn@doi [Annu. Rev. Fluid.
  Mech.] {https://doi.org/10.1146/annurev-fluid-010814-014556}, 47, 163

\bibitem[\protect\citeauthoryear{Liot et~al.,}{Liot
  et~al.}{2016}]{liot2016simultaneous}
Liot O.,  et~al., 2016, \mn@doi [J. Fluid Mech.]
  {https://doi.org/10.1017/jfm.2016.190}, 794, 655

\bibitem[\protect\citeauthoryear{Long, Mound, Davies  \& Tobias}{Long
  et~al.}{2020}]{long2020scaling}
Long R.~S.,  Mound J.~E.,  Davies C.~J.,   Tobias S.~M.,  2020, \mn@doi [J.
  Fluid Mech.] {https://doi.org/10.1017/jfm.2020.67}, 889, A7

\bibitem[\protect\citeauthoryear{Lurie et~al.}{Lurie
  et~al.}{2017}]{lurie2017tidal}
Lurie J.~C.,  et~al., 2017, \mn@doi [\apj]
  {https://doi.org/10.3847/1538-3881/aa974d}, 154, 250

\bibitem[\protect\citeauthoryear{Mathis, Auclair-Desrotour, Guenel, Gallet  \&
  Le~Poncin-Lafitte}{Mathis et~al.}{2016}]{mathis2016impact}
Mathis S.,  Auclair-Desrotour P.,  Guenel M.,  Gallet F.,   Le~Poncin-Lafitte
  C.,  2016, \mn@doi [A\&A] {https://doi.org/10.1051/0004-6361/201527545}, 592,
  A33

\bibitem[\protect\citeauthoryear{Mazeh}{Mazeh}{2008}]{Mazeh2008}
Mazeh T.,  2008, \mn@doi [EAS] {https://doi.org/10.1051/eas:0829001}, 29, 1

\bibitem[\protect\citeauthoryear{Meibom \& Mathieu}{Meibom \&
  Mathieu}{2005}]{meibom2005robust}
Meibom S.,  Mathieu R.~D.,  2005, \mn@doi [\apj]
  {https://doi.org/10.1086/427082}, 620, 970

\bibitem[\protect\citeauthoryear{Meibom, Mathieu  \& Stassun}{Meibom
  et~al.}{2006}]{meibom2006observational}
Meibom S.,  Mathieu R.~D.,   Stassun K.~G.,  2006, \mn@doi [ApJ]
  {https://doi.org/10.1086/508252}, 653, 621

\bibitem[\protect\citeauthoryear{Monville, Vidal, C\'ebron  \&
  Schaeffer}{Monville et~al.}{2019}]{monville2019rotating}
Monville R.,  Vidal J.,  C\'ebron D.,   Schaeffer N.,  2019, \mn@doi [Geophys.
  J. Int.] {https://doi.org/10.1093/gji/ggz347}, 219, S195

\bibitem[\protect\citeauthoryear{Newton, Mondrik, Irwin, Winters  \&
  Charbonneau}{Newton et~al.}{2018}]{Newton2018}
Newton E.~R.,  Mondrik N.,  Irwin J.,  Winters J.~G.,   Charbonneau D.,  2018,
  \mn@doi [\apj] {https://doi.org/10.3847/1538-3881/aad73b}, 156, 217

\bibitem[\protect\citeauthoryear{Nielsen, Gizon, Schunker  \& Karoff}{Nielsen
  et~al.}{2013}]{nielsen2013rotation}
Nielsen M.~B.,  Gizon L.,  Schunker H.,   Karoff C.,  2013, \mn@doi [A\&A]
  {https://doi.org/10.1051/0004-6361/201321912}, 557, L10

\bibitem[\protect\citeauthoryear{Niemann, Kantz  \& Barkai}{Niemann
  et~al.}{2013}]{niemann2013fluctuations}
Niemann M.,  Kantz H.,   Barkai E.,  2013, \mn@doi [Phys. Rev. Lett.]
  {https://dx.doi.org/10.1103/PhysRevLett.110.140603}, 110, 140603

\bibitem[\protect\citeauthoryear{Ogilvie}{Ogilvie}{2014}]{ogilvie2014tidal}
Ogilvie G.~I.,  2014, \mn@doi [ARA\&A]
  {https://doi.org/10.1146/annurev-astro-081913-035941}, 52, 171

\bibitem[\protect\citeauthoryear{Ogilvie \& Lesur}{Ogilvie \&
  Lesur}{2012}]{ogilvie2012interaction}
Ogilvie G.~I.,  Lesur G.,  2012, \mn@doi [MNRAS]
  {https://doi.org/10.1111/j.1365-2966.2012.20630.x}, 422, 1975

\bibitem[\protect\citeauthoryear{{Ogilvie} \& {Lin}}{{Ogilvie} \&
  {Lin}}{2007}]{OL2007}
{Ogilvie} G.~I.,  {Lin} D.~N.~C.,  2007, \mn@doi [\apj]
  {https://doi.org/10.1086/515435}, 661, 1180

\bibitem[\protect\citeauthoryear{{Paxton}, {Bildsten}, {Dotter}, {Herwig},
  {Lesaffre}  \& {Timmes}}{{Paxton} et~al.}{2011}]{Paxton2011}
{Paxton} B.,  {Bildsten} L.,  {Dotter} A.,  {Herwig} F.,  {Lesaffre} P.,
  {Timmes} F.,  2011, \mn@doi [\apjs]
  {https://doi.org/10.1088/0067-0049/192/1/3}, 192, 3

\bibitem[\protect\citeauthoryear{{Paxton} et~al.,}{{Paxton}
  et~al.}{2013}]{Paxton2013}
{Paxton} B.,  et~al., 2013, \mn@doi [\apjs] {10.1088/0067-0049/208/1/4}, 208, 4

\bibitem[\protect\citeauthoryear{{Paxton} et~al.,}{{Paxton}
  et~al.}{2015}]{Paxton2015}
{Paxton} B.,  et~al., 2015, \mn@doi [\apjs]
  {https://doi.org/10.1088/0067-0049/220/1/15}, 220, 15

\bibitem[\protect\citeauthoryear{{Paxton} et~al.,}{{Paxton}
  et~al.}{2018}]{Paxton2018}
{Paxton} B.,  et~al., 2018, \mn@doi [\apjs]
  {https://doi.org/10.3847/1538-4365/aaa5a8}, 234, 34

\bibitem[\protect\citeauthoryear{{Paxton} et~al.,}{{Paxton}
  et~al.}{2019}]{Paxton2019}
{Paxton} B.,  et~al., 2019, \mn@doi [\apjs]
  {https://doi.org/10.3847/1538-4365/ab2241}, 243, 10

\bibitem[\protect\citeauthoryear{Penev, Sasselov, Robinson  \& Demarque}{Penev
  et~al.}{2007}]{penev2007dissipation}
Penev K.,  Sasselov D.,  Robinson F.,   Demarque P.,  2007, \mn@doi [\apj]
  {https://doi.org/10.1086/507937}, 655, 1166

\bibitem[\protect\citeauthoryear{Penev, Barranco  \& Sasselov}{Penev
  et~al.}{2009}]{penev2009direct}
Penev K.,  Barranco J.,   Sasselov D.,  2009, ApJ, 705, 285

\bibitem[\protect\citeauthoryear{Penev, Barranco  \& Sasselov}{Penev
  et~al.}{2011}]{penev2011three}
Penev K.,  Barranco J.,   Sasselov D.,  2011, \mn@doi [\apj]
  {https://doi.org/10.1088/0004-637X/734/2/118}, 734, 118

\bibitem[\protect\citeauthoryear{Pereira, Gissinger  \& Fauve}{Pereira
  et~al.}{2019}]{pereira2019noise}
Pereira M.,  Gissinger C.,   Fauve S.,  2019, \mn@doi [Phys. Rev. E]
  {https://dx.doi.org/10.1103/PhysRevE.99.023106}, 99, 023106

\bibitem[\protect\citeauthoryear{Price-Whelan \& Goodman}{Price-Whelan \&
  Goodman}{2018}]{price2018binary}
Price-Whelan A.~M.,  Goodman J.,  2018, \mn@doi [ApJ]
  {https://doi.org/10.3847/1538-4357/aae264}, 867, 5

\bibitem[\protect\citeauthoryear{Rasio, Tout, Lubow  \& Livio}{Rasio
  et~al.}{1996}]{Rasio1996}
Rasio F.~A.,  Tout C.~A.,  Lubow S.~H.,   Livio M.,  1996, \mn@doi [\apj]
  {10.1086/177941}, 470, 1187

\bibitem[\protect\citeauthoryear{Rieutord}{Rieutord}{2014}]{rieutord2014fluid}
Rieutord M.,  2014, Fluid {Dynamics}: {An} {Introduction}.
Springer, Berlin

\bibitem[\protect\citeauthoryear{Samadi, Goupil  \& Lebreton}{Samadi
  et~al.}{2001}]{Samadi2001}
Samadi R.,  Goupil M.-J.,   Lebreton Y.,  2001, \mn@doi [\aap]
  {https://doi.org/10.1051/0004-6361:20010213}, 370, 147

\bibitem[\protect\citeauthoryear{Schaeffer}{Schaeffer}{2013}]{schaeffer2013efficient}
Schaeffer N.,  2013, \mn@doi [Geochem. Geophys. Geosyst.]
  {https://doi.org/10.1002/ggge.20071}, 14, 751

\bibitem[\protect\citeauthoryear{Spiegel}{Spiegel}{1971}]{Spiegel1971}
Spiegel E.~A.,  1971, \mn@doi [\araa]
  {https://doi.org/10.1146/annurev.aa.09.090171.001543}, 9, 323

\bibitem[\protect\citeauthoryear{Terquem, Papaloizou, Nelson  \& Lin}{Terquem
  et~al.}{1998}]{terquem1998tidal}
Terquem C.,  Papaloizou J.,  Nelson R.,   Lin D.,  1998, \mn@doi [\apj]
  {https://doi.org/10.1086/305927}, 502, 788

\bibitem[\protect\citeauthoryear{Triaud et~al.}{Triaud
  et~al.}{2017}]{triaud2017eblm}
Triaud A. H. M.~J.,  et~al., 2017, \mn@doi [A\&A]
  {https://doi.org/10.1051/0004-6361/201730993}, 608, A129

\bibitem[\protect\citeauthoryear{Van~Eylen, Winn  \& Albrecht}{Van~Eylen
  et~al.}{2016}]{van2016orbital}
Van~Eylen V.,  Winn J.~N.,   Albrecht S.,  2016, \mn@doi [\apj]
  {https://doi.org/10.3847/0004-637X/824/1/15}, 824, 15

\bibitem[\protect\citeauthoryear{Verbunt \& Phinney}{Verbunt \&
  Phinney}{1995}]{verbunt1995tidal}
Verbunt F.,  Phinney E.~S.,  1995, A\&A, 296, 709

\bibitem[\protect\citeauthoryear{Vick \& Lai}{Vick \&
  Lai}{2020}]{vick2020tidal}
Vick M.,  Lai D.,  2020, \mn@doi [\mnras]
  {https://doi.org/10.1093/mnras/staa1784}, 496, 3767

\bibitem[\protect\citeauthoryear{Vidal \& Barker}{Vidal \&
  Barker}{2020}]{vidal2020turbulent}
Vidal J.,  Barker A.~J.,  2020, \mn@doi [\apjl]
  {https://doi.org/10.3847/2041-8213/ab6219}, 888, L31

\bibitem[\protect\citeauthoryear{Vidal \& C{\'e}bron}{Vidal \&
  C{\'e}bron}{2017}]{vidal2017inviscid}
Vidal J.,  C{\'e}bron D.,  2017, \mn@doi [J. Fluid Mech.]
  {https://doi.org/10.1017/jfm.2017.689}, 833, 469

\bibitem[\protect\citeauthoryear{Vidal \& Schaeffer}{Vidal \&
  Schaeffer}{2015}]{vidal2015quasi}
Vidal J.,  Schaeffer N.,  2015, \mn@doi [Geophys. J. Int.]
  {https://doi.org/10.1093/gji/ggv282}, 202, 2182

\bibitem[\protect\citeauthoryear{Vidal, C{\'e}bron, Schaeffer  \&
  Hollerbach}{Vidal et~al.}{2018}]{vidal2018magnetic}
Vidal J.,  C{\'e}bron D.,  Schaeffer N.,   Hollerbach R.,  2018, \mn@doi
  [MNRAS] {https://doi.org/10.1093/mnras/sty080}, 475, 4579

\bibitem[\protect\citeauthoryear{Vidal, C{\'e}bron, ud Doula  \& Alecian}{Vidal
  et~al.}{2019}]{vidal2019binaries}
Vidal J.,  C{\'e}bron D.,  ud Doula A.,   Alecian E.,  2019, \mn@doi [\aap]
  {https://doi.org/10.1051/0004-6361/201935658}, 629, A142

\bibitem[\protect\citeauthoryear{Zahn}{Zahn}{1966}]{zahn1966marees}
Zahn J.-P.,  1966, Ann. Astrophys, 29, 489

\bibitem[\protect\citeauthoryear{Zahn}{Zahn}{1989}]{zahn1989tidal}
Zahn J.-P.,  1989, A\&A, 220, 112

\bibitem[\protect\citeauthoryear{Zahn}{Zahn}{2008}]{zahn2008tidal}
Zahn J.-P.,  2008, \mn@doi [EAS Publications Series]
  {https://doi.org/10.1051/eas:0829002}, 29, 67

\bibitem[\protect\citeauthoryear{Zahn \& Bouchet}{Zahn \&
  Bouchet}{1989}]{zahn1989tidalb}
Zahn J.-P.,  Bouchet L.,  1989, A\&A, 223, 112

\bibitem[\protect\citeauthoryear{von Boetticher et~al.}{von Boetticher
  et~al.}{2019}]{von2019eblm}
von Boetticher A.,  et~al., 2019, \mn@doi [A\&A]
  {https://doi.org/10.1051/0004-6361/201834539}, 625, A150

\makeatother
\end{thebibliography}


\appendix
\section{MESA Code Parameters}
\label{MESA}
We use MESA version 12778 \citep{Paxton2011, Paxton2013, Paxton2015, Paxton2018, Paxton2019}. The inlist file that we use is given below. We alter {\verb initial_mass } to generate a given stellar model.
\begin{verbatim}
&star_job
  create_pre_main_sequence_model = .true.
/ !End of star_job namelist
&controls
! starting specifications
    initial_mass = 1.0
    initial_z = 0.02d0
    MLT_option = 'Henyey'
    max_age = 5.0d10
    max_years_for_timestep = 1.0d8      
    use_dedt_form_of_energy_eqn = .true.
    use_gold_tolerances = .true.
    mesh_delta_coeff = 0.3
    when_to_stop_rtol = 1d-6
    when_to_stop_atol = 1d-6
/ !End of controls namelist
\end{verbatim}

\bsp	
\label{lastpage}
\end{document}